\documentclass{aa}

\usepackage{lineno}
\nolinenumbers
\usepackage[T1]{fontenc}
\usepackage{CJKutf8}
\usepackage{hyperref}
\usepackage{mathtools}
\usepackage{amsmath} 
\usepackage{physics}
\usepackage{amssymb}
\usepackage{enumerate}
\usepackage{bm}
\usepackage{graphicx}
\usepackage{subcaption}
\usepackage{soul}
\graphicspath{{./}{figures/}}

\begin{document}

\title{Magnetar Formation from Accretion Induced Collapse of White Dwarfs}

\author{
Lu\'is Felipe Longo Micchi\inst{1}
\fnmsep\thanks{ORCID: 0000-0002-7500-7384}
\email{luis.felipe.longo.micchi@uni-jena.de}
\and
Patrick Chi-Kit Cheong\inst{2}
\fnmsep\thanks{ORCID: 0000-0003-1449-3363}
\email{patrick.cheong@berkeley.edu}
\thanks{N3AS Postdoctoral Fellow}
\and
David Radice\inst{3,6,7}
\fnmsep\thanks{ORCID: 0000-0001-6982-1008}
}

\institute{
Theoretisch-Physikalisches Institut, Friedrich-Schiller-Universit\"at Jena,
07743 Jena, Germany
\and
Department of Physics, University of California, Berkeley,
Berkeley, CA 94720, USA
\and
Institute for Gravitation and the Cosmos,
The Pennsylvania State University,
University Park, PA 16802, USA
\and
Department of Physics \& Astronomy,
University of New Hampshire,
Durham, NH 03824, USA
\and
Center for Nonlinear Studies,
Los Alamos National Laboratory,
Los Alamos, NM 87545, USA
\and
Department of Physics,
The Pennsylvania State University,
University Park, PA 16802, USA
\and
Department of Astronomy \& Astrophysics,
The Pennsylvania State University,
University Park, PA 16802, USA
}

\abstract
{}
{
We aim to characterize the post-collapse evolution of accretion-induced collapse (AIC) remnants of rapidly rotating, magnetized white dwarfs, focusing on their rotational, magnetic, and thermal structure, as well as the development of instabilities and their energy content.
}
{
We perform nine axis-symmetric general-relativistic neutrino magnetohydrodynamic (MHD) simulations of collapsing, rapidly rotating, magnetized white dwarfs. The simulations follow the system from collapse through bounce and up to $\sim$1 s post-bounce. The simulations are performed by the conformally flat general relativistic neutrino MHD code \texttt{Gmunu}.
}
{
The collapse produces a rapidly rotating proto-magnetar surrounded by a persistent accretion disk lasting at least $\sim 1$ s after bounce. The remnant mass and spin span 1.15--1.45 $M_{\odot}$ and 2.9--4.9 kHz, respectively, with stronger initial magnetic fields generally leading to lower rotation rates. During the first $\sim 10$ ms, the proto-magnetar exhibits global oscillations that drive both gravitational-wave emission and coherent modulation of the poloidal magnetic field energy. The magnetic energy evolution, normalized to its bounce value, follows an approximately universal behavior across all models. The remnant interior remains strongly magnetized ($\gtrsim 10^{13}$ G) and hot ($\gtrsim 20$ MeV) up to 1 s after bounce, with maxima of both quantities co-located in the inner $\sim 10$ km. The magnetic field topology shows surface poloidal fields of ${\sim}10^{12}$ G and toroidal fields of ${\sim}10^{14}$ G, with strong toroidal components extending into the equatorial region. When the magnetic energy exceeds the rotational energy ($\sim 10^{52}$ erg), the remnant core becomes unstable, leading to episodic magnetic flux expulsion, mass ejection, and flare-like activity in which magnetic energy is released and thermalized in the surrounding material.
}
{}

\keywords{
stars: neutron --
magnetic fields --
gravitational waves --
accretion, accretion disks --
MHD
}

\maketitle
 \nolinenumbers

\section{\label{sec:intro}Introduction}

The accretion-induced collapse (AIC) of white dwarfs (WDs) constitutes a physically motivated channel for the production of rapidly rotating neutron stars, neutron-rich ejecta, and relativistic outflows. The observation of long-duration gamma-ray bursts (LGRBs) accompanied by kilonova (KN) emission, such as GRB 211211A~\citep{2022Natur.612..223R, 2022Natur.612..228T, 2022Natur.612..236M, 2022Natur.612..232Y} and GRB 230703A~\citep{2023GCN.33405....1F, 2023GCN.33407....1D, 2023GCN.33411....1D, 2023GCN.33419....1E, 2023GCN.33429....1B}, further highlights the relevance of this channel as a potential source of both r-process nucleosynthesis and LGRB jets \citep{Cheong_2025_jet,combi2025jet}. This interpretation is especially interesting for GRB 230703A, although some authors claim that this scenario is disfavored ~\citep{notMerger}. In addition, AIC nucleosynthetic yields and kilonova light curves are broadly consistent with observational constraints \citep{Pitik:2026bjm}. In these previous works, it was shown that magnetic fields and rotation play a major role in the AIC physics. For example, in \citet{Cheong_2025_jet} it was shown that highly magnetized models eject mass and angular momentum more efficiently. In addition, the outflow increases to a higher velocity, expanding faster than the time necessary for the neutrino capture by neutrons to occur, leading to low $Y_e$ ejected material. These features have important observational and nucleosynthetic consequences, as shown in \citet{Pitik:2026bjm}.
More recently, 3D simulations of magnetized AIC events were performed by \citet{combi2025jet}, and \citet{Kuroda:2025iyj}, reinforcing the understanding that highly magnetized collapsing WD can indeed form (mildly) relativistic jets.

\begin{table*}[ht]
\centering
\hspace*{-0.6cm}
\begin{tabular}{lccc c ccccc}
\hline\hline
Model & Grid Size & Max. Res. &  $B_{\rm pol.}$ & $B_{\rm tor.}$ & $\Omega_{PNS}$ [kHz] & $E_{kin}^{PNS}$[ergs]& $E_{B_{tor}}^{PNS}$[ergs] & $E_{B_{pol}}^{PNS}$[ergs]\\
\hline
Bt0p1e9a & $R,z \gtrsim 2000$ km & $\Delta x \approx 488 m$  & $1\times10^{9}$  & 0 & 4.22 & 2.89$\times10^{52}$&2.22$\times10^{48}$&1.39e$\times10^{42}$\\
Bt0p1e10 & $R,z \gtrsim 2000$ km & $\Delta x \approx 488 m$  & $1\times10^{10}$ & 0 & 3.99 & 2.90$\times10^{52}$&2.02$\times10^{50}$&1.46e$\times10^{44}$\\
Bt0p1e11 & $R,z \gtrsim 2000$ km & $\Delta x \approx 488 m$  & $1\times10^{11}$ & 0 & 3.99 & 2.66$\times10^{52}$&1.24$\times10^{52}$&1.55$\times10^{46}$\\
Bt0p1e12 & $R,z \gtrsim 2000$ km & $\Delta x \approx 488 m$  & $1\times10^{12}$ & 0 & 2.49 & 9.09$\times10^{51}$&3.03$\times10^{51}$&9.26$\times10^{48}$\\
\hline
Bt0p1e9b  & $R,|z| \gtrsim 3\times10^5$ km & $\Delta x \approx 286 m$  & $1\times10^{9}$  & 0 & 4.22 & 3.01$\times10^{52}$&2.66$\times10^{48}$&2.13$\times10^{42}$\\
Bt1e8p1e9 & $R,|z| \gtrsim  3\times10^5$ km & $\Delta x \approx 286 m$  & $1\times10^{9}$  & $1\times10^{8}$ & 4.71 & 3.01$\times10^{52}$&2.71$\times10^{48}$&2.09$\times10^{42}$\\
Bt5e8p1e9 & $R,|z| \gtrsim  3\times10^5$ km & $\Delta x \approx 286 m$  & $1\times10^{9}$  & $5\times10^{8}$ &  3.93 & 3.02$\times10^{52}$&2.83$\times10^{48}$&2.06$\times10^{42}$\\
Bt1e9p1e9 & $R,|z| \gtrsim  3\times10^5$ km & $\Delta x \approx 286 m$  & $1\times10^{9}$  & $1\times10^{9}$ & 4.11 & 3.01$\times10^{52}$&2.85$\times10^{48}$&2.16$\times10^{42}$\\
Bt2e9p1e9 & $R,|z| \gtrsim  3\times10^5$ km & $\Delta x \approx 286 m$  & $1\times10^{9}$  & $2\times10^{9}$ & 4.61 & 3.01$\times10^{52}$&2.84$\times10^{48}$&2.17$\times10^{42}$\\
\hline\hline
\end{tabular}
\caption{The main differences between our nine models are listed here. The magnetic field and its topology are varied by orders of magnitude to investigate their effect on the remnant. The different resolutions and grid sizes are also shown in the table. Remnant's features such as angular velocity ($\Omega^{\rm PNS}$), its total kinetic energy ($E_{kin}^{\rm PNS}$), and the magnetic energy of each of the field's components ($E_{B_{pol}}^{\rm PNS}$ and $E_{B_{tor}}^{\rm PNS}$) as extracted at 1 second. Here we assume the PNS to be the central region with $\rho>10^{11}$g/cm$^{3}$. The angular velocity is extracted at the equatorial plane at $r\sim5$km. Worth mentioning that the top four models are run with bitant symmetry, while the bottom five also include negative values of $z$.}
\label{tab:models}
\end{table*}

Given their accretion history, AIC are expected to be rapidly rotating~\citep{2003ApJ...583..885P, 2003ApJ...598.1229P, 2003ApJ...595.1094U, 2004ApJ...615..444S, 2018ApJ...869..140K}, which impose constraints on our expectations on the angular momentum of these events.
On the other hand, the strength of the magnetic field involved in such events is not well-known. While the highest magnetic field ever detected in association with a WD is of order $10^{9}$ Gauss \citep{2021Natur.595...39C}, this object represents a stable and isolated WD not corresponding to the conditions faced by a WD on the verge of collapse. Previous works have shown that WDs can face much greater magnetic fields if they go through a merger or accretion history \citep{2015ApJ...806L...1Z, 2024arXiv240702566P,2015SSRv..191..111F, 2020IAUS..357...60K, 2020AdSpR..66.1025F, 2015ApJ...806L...1Z, 2024A&A...691A.179P}. The collapse of a WD above the Chandrasekhar mass limit occurs on a timescale of ${\sim}100$ milliseconds, rendering direct detection of the collapse unlikely. However, such systems may still be observable prior to collapse, while they remain thermally stable and luminous. We argue that the structure of the long-lived remnant of accretion-induced collapse (AIC) is therefore key to inferring progenitor WD properties.

Although previous works analyzed the impact of the WD progenitor on the ejecta \citep{Batziou_2025}, we note a lack of investigation of the final PNS's properties in the literature.
In this work, we present a systematic investigation of the remnant left behind by an explosive AIC event, with a particular focus on the influence of different topologies of the initial WD's magnetic field. Here, we made use of the conformally flat general relativistic neutrino magnetohydrodynamic (GR$\nu$MHD) code \texttt{Gmunu}~\citep{2020CQGra..37n5015C, 2021MNRAS.508.2279C, 2022ApJS..261...22C, 2023ApJS..267...38C, 2024ApJS..272....9N}. The paper is organized as follows. In Section \ref{sec:methods} and \ref{sec:ID}, we review the simulation methods and initial data profiles used for our models, respectively.
Section \ref{sec:results} displays our results regarding the remnant properties, and a discussion of the formation of magnetic flares. Our final conclusions and discussions can be found in Section \ref{sec:conclusions}. In Appendix \ref{sec:Diagnostics}, we show some magnetic-field diagnostic quantities of our simulations.

\section{\label{sec:methods}Methods }

This work follows the same methodology as \citet{Cheong_2025_jet}. 
In particular, we use the GR$\nu$MHD code \texttt{Gmunu}~\citep{2020CQGra..37n5015C, 2021MNRAS.508.2279C, 2022ApJS..261...22C, 2023ApJS..267...38C, 2024ApJS..272....9N} to solve the conformally flat GRMHD Einstein field equations, as well as the gray M1 neutrino radiation transport scheme from \cite{2022MNRAS.512.1499R}. The neutrino rates are the same as the ones available in the \texttt{WhiskyTHC} code~\citep{2022MNRAS.512.1499R}. The M1 scheme is only used after the core-bounce time, while the effective de-leptonization scheme of \cite{2005ApJ...633.1042L} is used at earlier times. While the finite temperature equation-of-state (EOS) LS220~\citep{1991NuPhA.535..331L} is chosen for all our models, at low densities we assume an ideal gas of adiabatic index $\Gamma=5/3$ plus radiation terms, see appendix A of \citet{Cheong_2025_jet}. Appendix B of the aforementioned paper describes the ejecta extraction method used for our diagnostics. The Riemann solver (HLL), reconstruction method (PPM), and time integrator (IMEXCB3a) used are described in ~\citet{harten1983upstream},~\citet{1984JCoPh..54..174C}, and ~\citet{2015JCoPh.286..172C}, respectively.

\begin{figure*}
    \centering
    \includegraphics[width=0.9\linewidth]{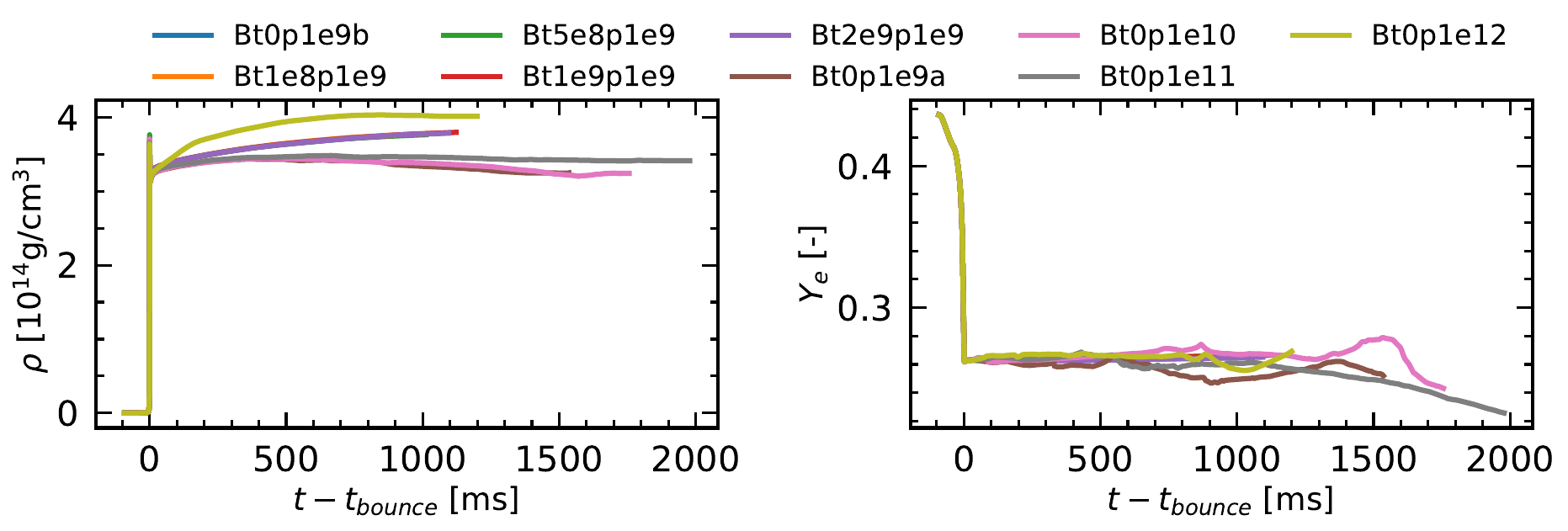}
    \caption{Here we show the central density (estimated by the position of maximum density) and the corresponding fluid's electron fraction. Note that the time axis has been translated by the reference value of $t_{bounce}$, all our figures should follow this standard. Note that the magnetic field of different strengths does not alter the collapse phase of the AIC significantly. }
    \label{fig:Center}
\end{figure*}

\section{\label{sec:ID} Models and Initial Data}

In this work, we extend the initial data set described in \citet{Cheong_2025_jet} by adding five other models of varying magnetic field topology. In total, we summarize nine simulations and their main differences in Table \ref{tab:models}. 
Apart from the main difference in the topology of the magnetic field, all models share the same physical quantities. Adopting the LS220 equation-of-state (EOS;\citealt{1991NuPhA.535..331L}), we use the \texttt{RNS} code~\citep{1995ApJ...444..306S} to generate our initial data. Setting the central energy density to $\epsilon_{c}/c^2 = 1\times10^{10}~\rm{g \cdot cm^{-3}}$~\citep{1991ApJ...367L..19N, 2004A&A...419..623Y, 2005A&A...435..967Y, 2006ApJ...644.1063D, 2007ApJ...669..585D} our WDs of symmetric composition ($Y_{e}=0.5$) are found to have a mass of $1.5\,M_{\odot}$. We also constructed our initial data with rapid rotation ($\Omega \approx 5~{\rm Hz}$), leading to an oblate star of $a_{r}=0.75$, where $a_{r}$ is defined as the ratio between the polar and equatorial radii. This choice represents ${\sim} 80\%$ of the Keplerian limit.
Rapid rotation is an expected feature, as AIC progenitors have a history of accretion ~\citep{2003ApJ...583..885P, 2003ApJ...598.1229P, 2003ApJ...595.1094U, 2004ApJ...615..444S, 2018ApJ...869..140K}. The temperature profile of our models follows the prescription of \cite{2006ApJ...644.1063D, 2007ApJ...669..585D}, implying that
\begin{equation}
    T= T_{c} \left(\dfrac{\rho_{c}}{\rho}\right)^{0.35}
\end{equation}
where $T_{c}=5\times10^{9}K$ is the central temperature and $\rho_{c}$ is the central density.

Regarding magnetic fields, they are described by the vector potential \citep[e.g.,][]{2007PASJ...59..771S, 2021MNRAS.508.6033V}: 
\begin{equation}
	\left(A^{\hat{r}}, A^{\hat{\theta}}, A^{\hat{\phi}}\right) = \frac{r_0^3 r}{2\left(r^3+r_0^3\right)}\left(B_{\rm tor} \cos\theta, 0, B_{\rm pol}  \sin\theta \right),
\end{equation}
where $r$ is the distance from the center and $r_{0}$ is set to 600 km~\citep{2007ApJ...669..585D}. For all of our models, the magnetic field's symmetry axis is aligned with the angular momentum of the progenitor.


\section{\label{sec:results}Results}

\subsection{AIC Dynamics} 

A WD that accretes enough mass to surpass the total mass of $\approx 1.4 M_{\odot}$, a value known as the Chandrasekhar mass limit \citep{ChandraMass}, becomes unstable, as the gravitational pull is stronger than what the electron degeneracy pressure can support. In this scenario, the collapsing WD will have two possible fates: during the collapse, a carbon ignition takes place, and if sustained, the flame can generate a runaway process causing the thermonuclear explosion known as Supernova type Ia (SNIa); or the infall becomes supersonic before the burning takes place, making it impossible for the explosion to stop the collapse. The second scenario gives rise to the AIC event. The exact mechanism to determine if the thermonuclear explosion ignites or nor is still an open question that dates back to \citet{1991ApJ...367L..19N}, where it was argued that heavy ($M>1.1 M_{\odot}$) O+Ne+Mg WDs undergoing slow accretion rates ($\lesssim 10^{8} M_{\odot}/s$) are ideal candidates for AICs. For a broader discussion on the topic, we direct the readers to reviews on the topic \citet{Niemeyer_1997} and \citet{RuiterSeitenzahl2025}

In this work, we remain agnostic with respect to the accretion history of the progenitor star, starting the simulation with an isolated star that is forced to collapse by imposing an electron capture process by parameterizing the electron fraction ($Y_{e}$) as a function of density ($Y_{e}(\rho)$). Figure \ref{fig:Center} shows the evolution of the central density and central $Y_{e}$ of our models as a function of time. As the central density of the WD increases, the electron fraction diminishes, essentially shutting off the electron degeneracy pressure that provided support against gravitational collapse to the WD.
As the star collapses, its central density reaches values of $\gtrsim 3 \times 10^{14}$ g/cm$^{3}$ in ${\sim} 100$ ms after the start of the simulation. This short period of collapse comes to an end due to the stiffness reached by the core of the PNS at nuclear saturation densities. At this moment, the core goes through a sudden relaxation, period shown by the decay of the central density, in a process referred to as core bounce. At bounce time ($t=t_{bounce}$), kinetic energy of the order ${\sim}10^{50}$ ergs is injected into the surrounding material, giving rise to the explosive event called AIC. Figure \ref{fig:Center} shows that all our models collapse and bounce in similar time scales, indicating a very low influence of the magnetic field strength on these time scales.

\begin{figure*}
    \centering
    \includegraphics[width=1.0\linewidth]{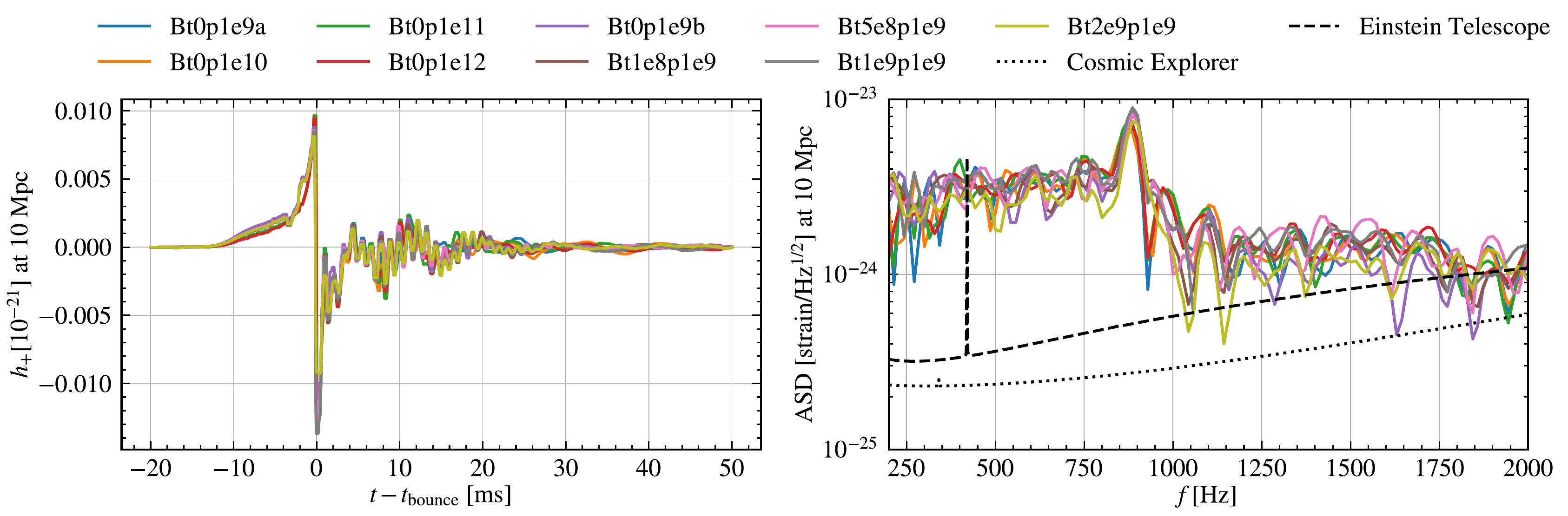}
    \caption{Gravitational wave signal recovered from our models by means of the moment of inertia, see Equation \ref{eq:GW}. All our models show a prompt signal following the bounce, mainly containing $900$Hz radiation. The lack of a late time emission is due to the axisymmetry of our models, which prevent the $m=1$ instability found in \citet{2023MNRAS.525.6359L}. Our findings support the expectation of detections of such events by CE and ET up to $\sim$10Mpc.   }
    \label{fig:GW}
\end{figure*}

\begin{figure*}
    \centering
    \includegraphics[width=1.0\linewidth]{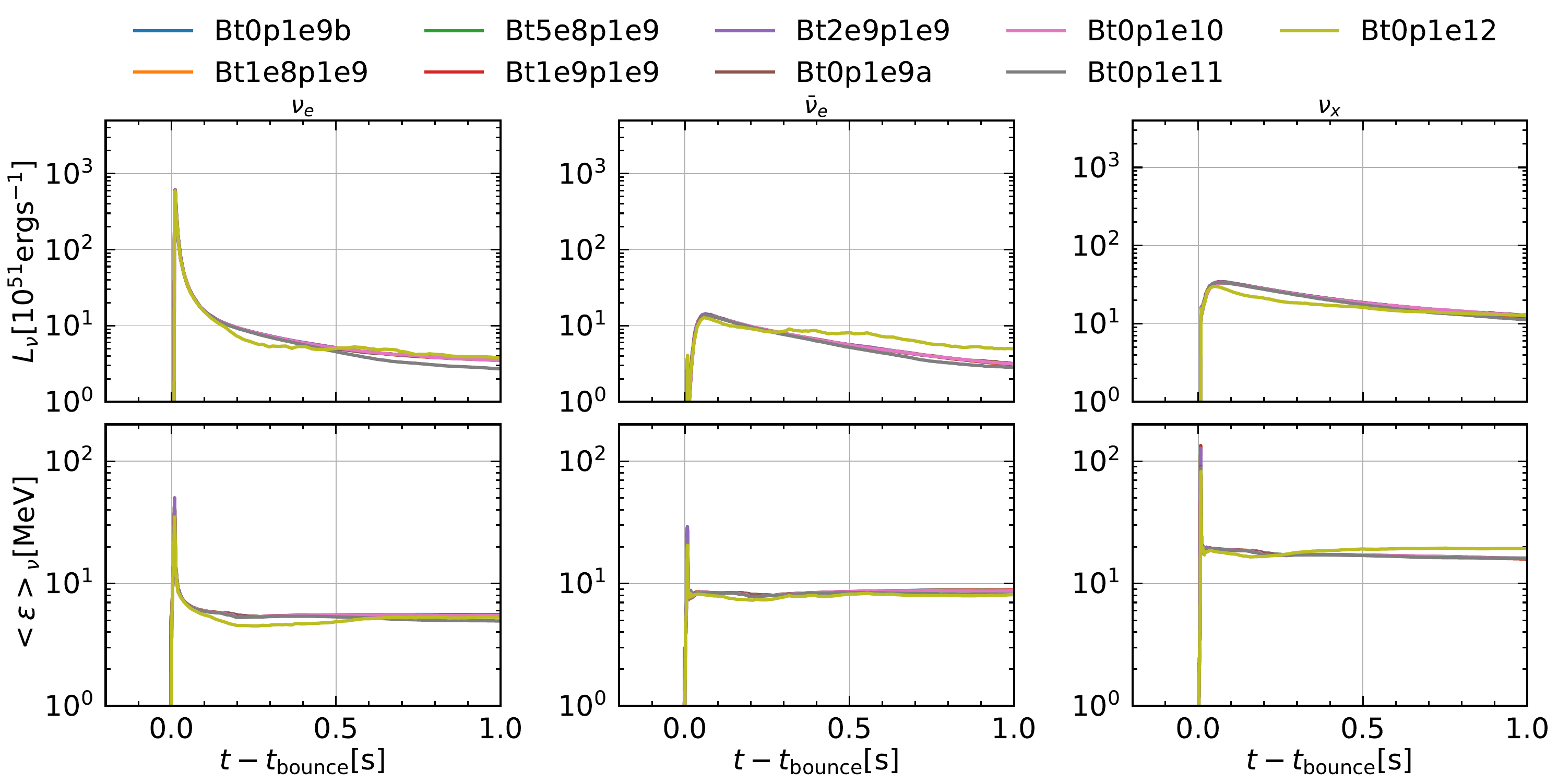}
    \caption{Total neutrino luminosities and average energies as a function of time for each one of the neutrino flavors. We found that a higher magnetic field leads to a higher asymptotic value of $L_{\bar{\nu}_{e}}$ with no significant increase in the average neutrino energy. }
    \label{fig:Nu}
\end{figure*}

\subsection{Gravitational Wave Signals}

In Figure~\ref{fig:GW}, we show the $h_{+}$ mode of the expected GW signals of all our models.
All our waveforms are extracted in the equatorial plane via the quadrupole formula, which reads: 
\begin{equation}\label{eq:GW}
    h_{+} = - \dfrac{1}{r} \dfrac{GM}{c^{4}}  \left( \ddot{I}_{RR} - 2 \ddot{I}_{zz} \right),
\end{equation}
where $R$ is the spherical radius.
Despite their very different magnetic field configurations, all models show very similar strains and frequency content. The similarity between the gravitational radiation signals signals that the magnetic field is not the dominant factor driving the fluid's motion, at least for $t-t_{\rm bounce}\lesssim 20$ms. The absence of GW signals for $t-t_{\rm bounce}\gtrsim 50$ms is a reflection of the axisymmetry of our models. This prevents us from studying the presence of the $m=1$ instability found in \citet{2023MNRAS.525.6359L}. The frequency content of our models peaks at $f\sim900$Hz, similar to the kilohertz content found for 3D GRHD simulations \citep{2023MNRAS.525.6359L} and to the 700-800 Hz found in former axysymmetric conformally flat simulations of \citet{2010PhRvD..81d4012A}.

In the right panel of Figure~\ref{fig:GW}, we compare the signals' power spectrum against the noise budget of future GW detectors, such as Einstein Telescope (ET) and Cosmic Explorer (CE). Our results show that such events are expected to be well within their detection capabilities if they take place within our own local group of galaxies ($\sim$10~Mpc).

In our analysis, we find no obvious imprint of magnetic field strength in the GW sector, as all our predicted signals are robust when varying the progenitor's magnetic field by 3 orders of magnitude. Due to the axis-symmetry of our models, the $m=1$ instability that leads to late-time GW emission \citep{2023MNRAS.525.6359L} is not excited, therefore our estimates of detectability should be regarded as lower bounds.

\subsection{Neutrino Radiation}

In the top panels of Figure~\ref{fig:Nu}, we compare the neutrino radiation of all our models. Regarding the total luminosity of different neutrino flavors, we notice that the Bt0p1e12 model stands out from the others, especially with respect to $\bar{\nu}_{e}$. The late-time behavior of this model emits higher energy fluxes in this channel. This is related to the lower $Y_{e}$ material expelled by this model, see \citet{Cheong_2025_jet}. This electron-rich material presents more favorable conditions for the inverse $\beta$-decay process, increasing the overall production of $\bar{\nu}_{e}$. This is in alignment with the bottom panels of Figure~\ref{fig:Nu}, which shows that the average anti-electron neutrino energy among the different models is not drastically affected by increasing magnetic fields. 

Previous GRHD 3D studies with similar initial WD profiles \citep{2023MNRAS.525.6359L} found neutrino luminosities (average energies) to be ${\sim}3\times10^{52}, {\sim}2\times10^{52}$, and ${\sim}2\times10^{52}$erg/s ($20, 8$, and $10$ MeV) for $\nu_{x},\nu_{e}$, and $\bar{\nu}_{e}$, respectively. In this work, we report these values to be $2{-}3 \times 10^{52}, 0.5{-}0.8 \times 10^{52}$, and $0.3{-}0.5 \times 10^{52}$erg/s (${\sim}20,\ {\sim}5$, and ${\sim}9$ MeV) in the same order. The biggest difference is seen in $\nu_{e}$ luminosity. This factor of ${\sim} 4$ is easily explained by the longer time evolution performed in the current work and the monotonically decaying behavior of $L_{\nu_{e}}$ in the post-bounce phase of the AIC event.
We also note that the highly magnetized model neutrino luminosities display slightly different behaviors when compared to the other models. For example, $L_{\bar{\nu}_{e}}$ is $\sim$2 times brighter for Bt0p1e12 at 1 second after bounce. In fact, all neutrinos' luminosities of Bt0p1e12 seem to deviate from the general trend for $t-t_{\rm bounce}\sim 200$ms. This will later be connected to a short period of instability faced by this model. 

\begin{figure}
    \centering
    \includegraphics[width=\linewidth]{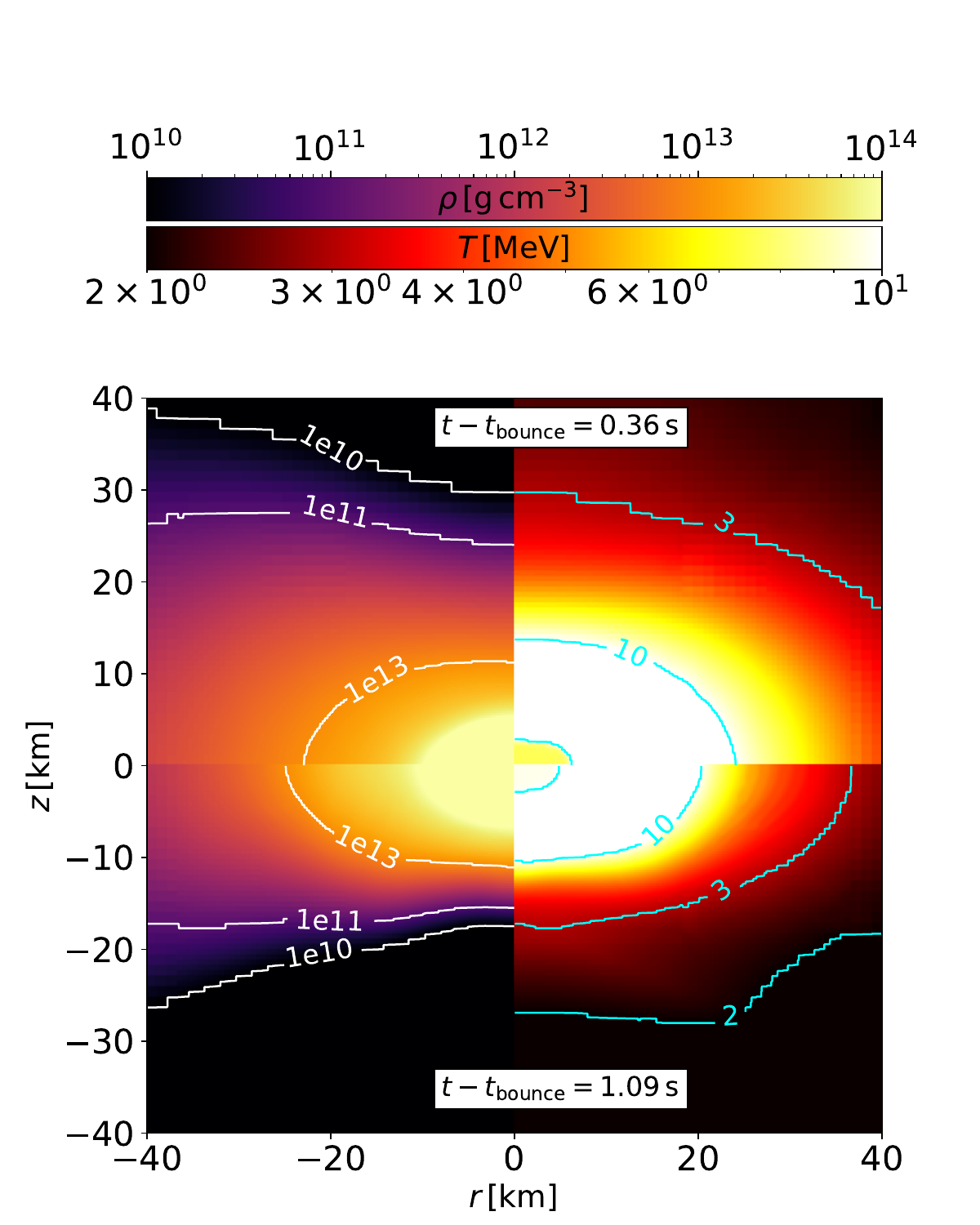}

    \caption{Temperature and density profiles for model Bt2e9p1e9. Here we see that the regions where $\nu_{x}$ become untrapped ($\rho \sim 10^{13}$ g/cm$^3$) is very close to the PNS' core and present temperatures $\gtrsim10$ Mev. On the other hand, $\rho \sim 10^{11}$ g/cm$^3$ regions are related to the still existing accretion disk, which displays temperatures $\lesssim 10$MeV. This explains the factor 2 connecting the average neutrino energies of heavy lepton neutrino and (anti-)electron neutrinos found in Figure~\ref{fig:Nu}. We find that the PNS is extremely hot ($\gtrsim 10$ MeV) when compared to an isolated NS (typically $\sim$ MeV). We note that the temperature peaks off-center of the star even for 1 second after the bounce. }
    \label{fig:NeutrinoSpheres}
\end{figure}

Regarding the neutrino average energies, one may note the systematically larger $\nu_{x}$ average energy compared to other neutrino flavors. We trace this difference to the fact that, while $\nu_{x}$ cease to be trapped at regions where $\rho\sim 10^{13}$ g/cm$^3$, $\nu_{e}$ and $\bar{\nu}_{e}$ are released where $\rho\sim 10^{11}$ g/cm$^3$. Figure~\ref{fig:NeutrinoSpheres} shows that $\rho\sim 10^{13}$ are very central to the PNS, but $\rho\sim 10^{11}$ are well localized with the still existing accretion disk, where temperatures are systematically lower. This explains the higher average energy found for heavy-lepton neutrinos. Figure~\ref{fig:NeutrinoSpheres} shows one representative model, but the qualitative behavior described is seen for all models. The neutrinos' average energies are more robust than their luminosities when varying the initial magnetic field. In particular, all of our models show the same energy hierarchy at all times.

Our studies suggest that, similarly to the results found in \citet{2023MNRAS.525.6359L}, the neutrino's luminosities are of the order of $10^{52}$ erg/s. 
In \citet{Batziou_2025}, where several ${\sim} 7$ second-long 2D general-relativistic hydrodynamical simulations were presented, the neutrinos' luminosities were found to vary between $10^{51}$ and $10^{52}$ erg/s at 1 second. The larger variance of the neutrinos' properties for this previous work is due to the larger variation of the initial profile of the progenitor WD.
Regarding their average energy, the heavy-lepton neutrinos have an average energy that is typically (2)4 times greater than the (anti-)electron neutrino, achieving an average of 20 MeV per neutrino. Similarly to this previous study, we found that the different flavors of neutrinos become untrapped at different temperatures. According to Figure~\ref{fig:NeutrinoSpheres}, while $\nu_{x}$ are released in the inner PNS regions where $T\gtrsim10$~MeV, $\nu_{e}$ and $\bar{\nu}_{e}$ are released from the extended disk where $T\gtrsim5$~MeV.
The previous work of \citet{Batziou_2025} found neutrinos' average energies varying between 6-15 MeV at ${\sim} 1$ second. Although our results show a fixed hierarchy of $\langle \epsilon_{\nu_x} \rangle \;>\; \langle \epsilon_{\nu_e} \rangle \;>\; \langle \epsilon_{\bar{\nu}_e} \rangle$, the results of this previous work found no such global ordering, the hierarchy being model dependent.
Another difference between our neutrinos' average energies is the secular increase seen for $\langle \epsilon \rangle_{\nu_{e}}$ found by \citet{Batziou_2025}. The authors blame this behavior on the evolution of the neutrino spheres that move inward as the star cools. We understand that our simulations may be too short to capture this feature.
It is worth mentioning that while our work relies on a gray neutrino scheme, some of the aforementioned works have multi-group transport and a more complete set of neutrino reaction rates. The different neutrino treatment may be the reason for the seen discrepancies.

\subsection{Remnant }

\begin{figure*}
    \centering
    \includegraphics[width=1.0\linewidth]{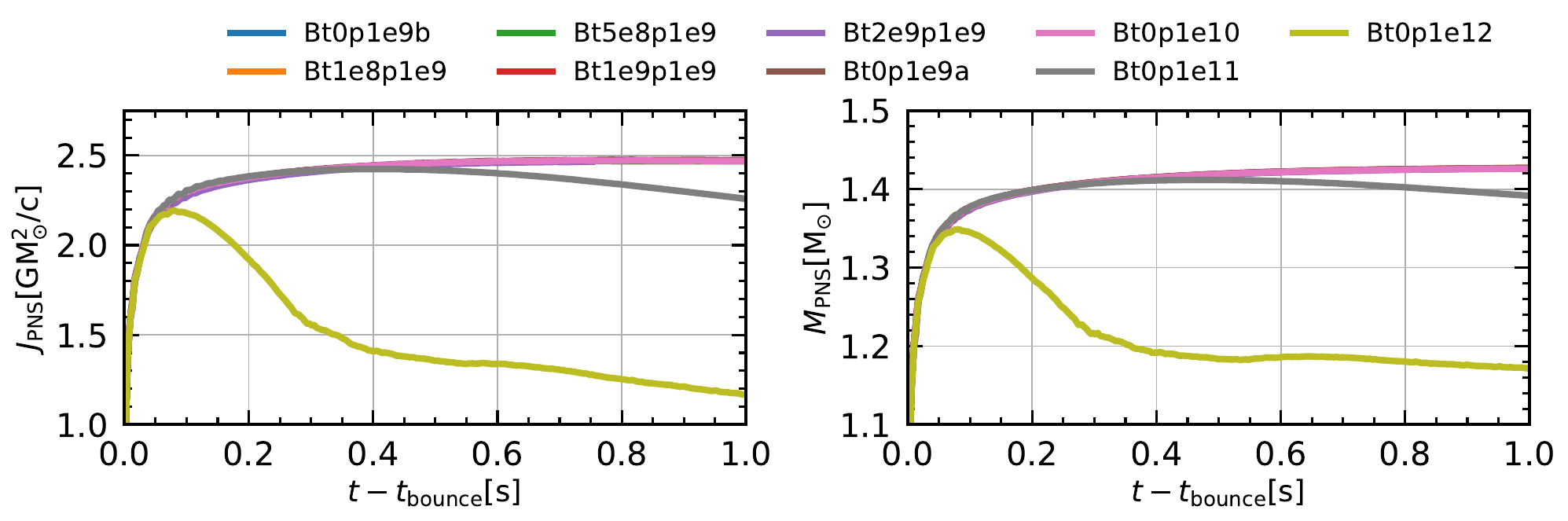}
    \caption{ Angular momentum and mass of the PNS as a function of time. We define the PNS as the region with $\rho \geq 10^{11}$g/cm$^{3}$. We note that apart from the Bt0p1e12, all models produce similar PNS with respect to their final mass (${\sim} 1.4 M_{\odot}$) and angular momentum ($J\sim2.4 G M_{\odot}^{2}/c$). We attribute the big difference encountered by the Bt0p1e12 model to the stronger magnetic wind faced by this model.}
    \label{fig:RemnantJM}
\end{figure*}

\begin{figure*}
    \centering
    \includegraphics[width=1.0\linewidth]{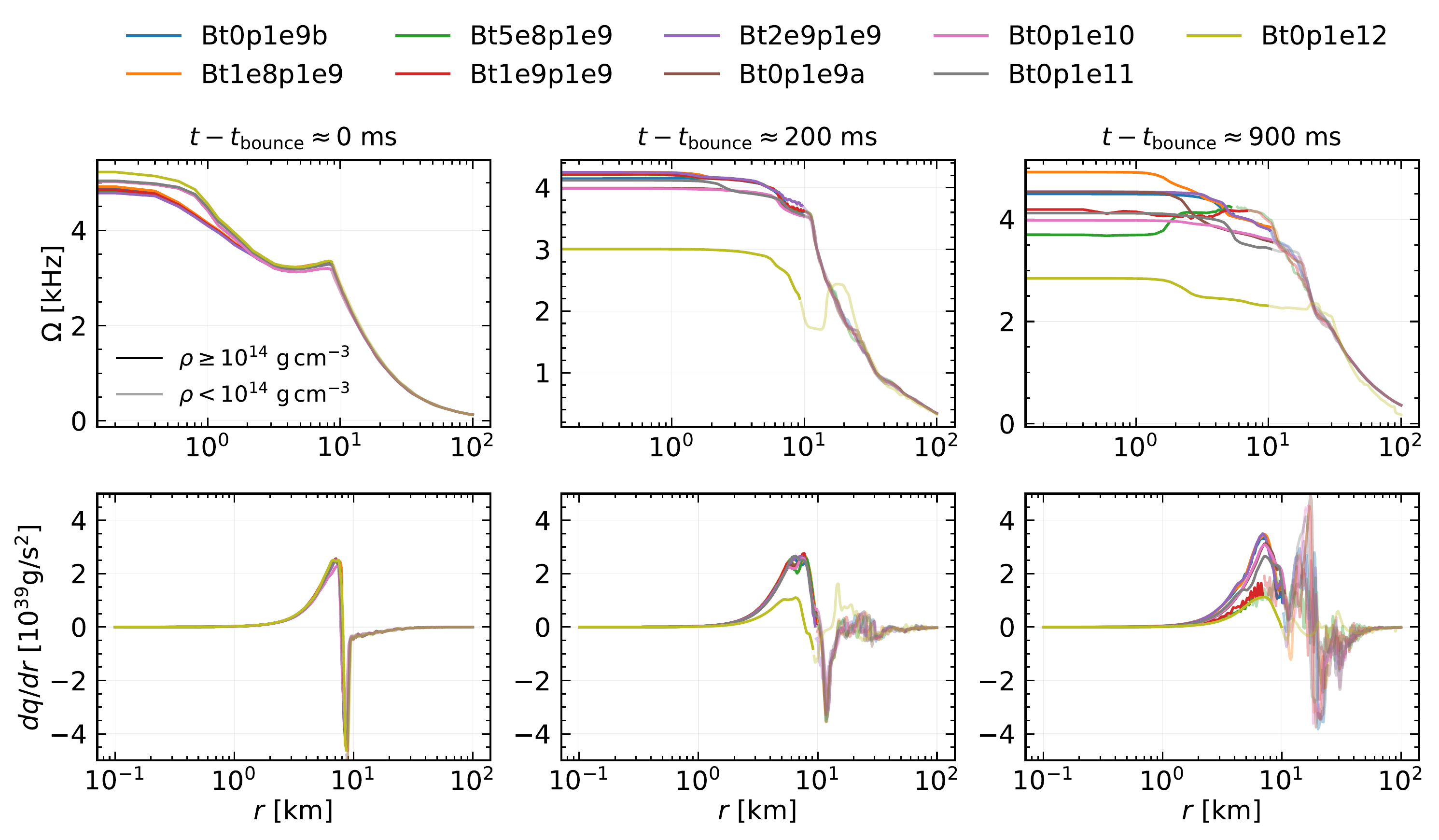}
    \caption{Radial (z=0) rotational profiles of our models at the equatorial plane for 3 different times. At bounce times, all of our models display a differential rotation profile. While most of our models achieve a solid rotation in the PNS core, Bt0p1e12 develops a Rayleigh-Taylor instability due to the depression in its angular velocity ($\Omega$) profile. This also appears as a negative derivative of $q$($\equiv \rho (\Omega R^{2})^{2}$), see more details in Figure~\ref{fig:Rotation2}. }
    \label{fig:Rotation1}
\end{figure*}

\begin{figure*}
    \centering
    \includegraphics[width=1.0\linewidth]{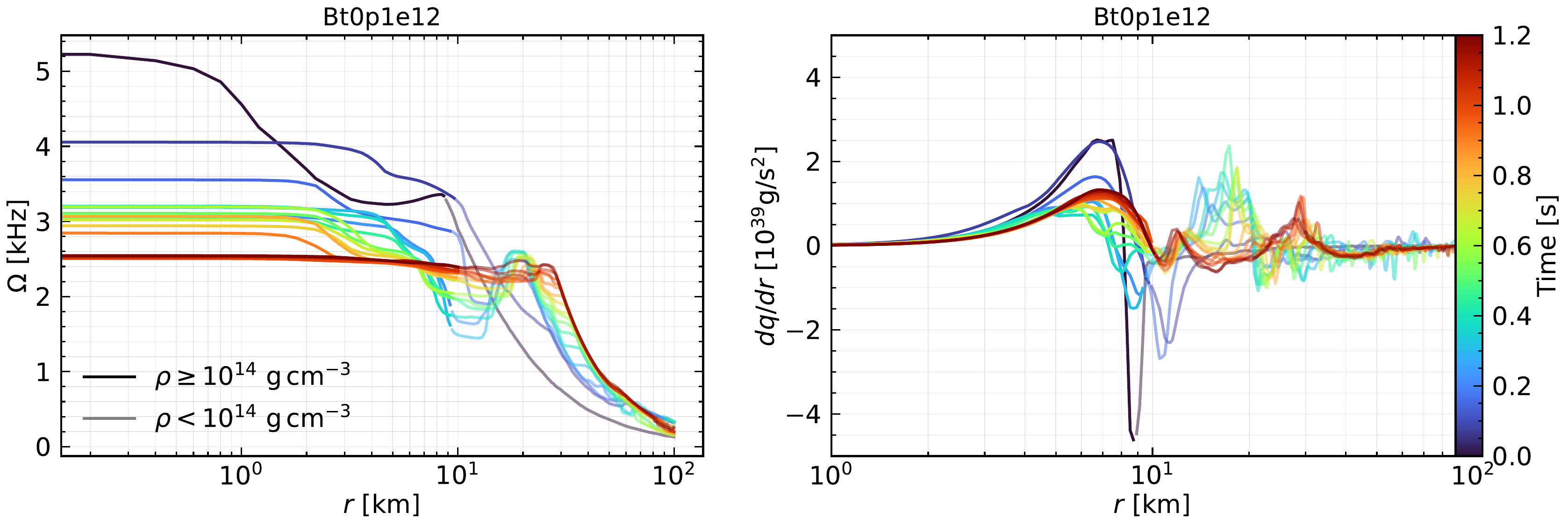}
    \caption{Radial (at z=0) rotational profile of Bt0p1e12 as a function of time. Here we see in more detail that the PNS’ core transitions between a differentially rotating profile to a solid rotation. We note that the transition is not without temporarily ($\sim$0.2-0.6 seconds after bounce) satisfying the Rayleigh instability criteria ($dq/dr < 0$). Here we see that such instability lasts for $\sim200$ ms before being quenched. This instability is related to convective motions in the outer layers of the PNS.  }
    \label{fig:Rotation2}
\end{figure*}

In Figure \ref{fig:RemnantJM} we show the PNS angular momentum ($J_{\rm PNS}$) and mass ($M_{\rm PNS}$). We find that, largely independently of the magnetic-field topology, the remnants evolve toward a mass of ${\sim} 1.4~M_{\odot}$ and a spin of ${\sim} 2.3~GM_{\odot}/c^{2}$. For most models, the PNS continues to accrete matter throughout the simulated interval, leading to a sustained increase in both mass and angular momentum up to at least $\gtrsim 1$ s after bounce. Consistent with this behavior, $\dot{J}$ remains positive whenever $\dot{M}>0$, indicating that ongoing accretion efficiently replenishes the angular momentum of the remnant and suppresses net spin-down. 
Only the models with the strongest initial poloidal magnetic fields ($\gtrsim 10^{11}$ G) depart significantly from these general trends. In these cases, particularly around $t-t_{\rm bounce}\sim0.2$ s, the PNS is no longer accreting but actually ejecting matter. As a result, the models Bt0p1e11 and Bt0p1e12 are the only cases that exhibit an early transition to spin-down, beginning at approximately 0.5 s and 0.1 s after bounce, respectively.

Figure \ref{fig:Rotation1} shows the angular velocity profiles along the equatorial plane as a function of the radius. It’s possible to see that, although all the profiles show a differential rotation at the time of bounce, most models’ PNS cores approach a solid rotation profile with $\Omega\sim3-5\times10^{3}$~Hz. A notable exception is Bt5e8p1e9, whose core seems to be very dynamic for the time of our simulations. The effect of the systematic increase of the magnetic field strength is non-monotonic, but as a general rule, more magnetized models achieve core solid rotation earlier. The Bt0p1e12 model has the lowest angular velocity at $\sim$1 second after bounce, even though its rotational profile is very similar to other models at the bounce time. The lower panels of Figure~\ref{fig:Rotation1} show the radial derivative of
\begin{equation}
    q \equiv \rho j^{2} = \rho (\Omega R^{2})^{2},
\end{equation}
where $j$ is the specific angular momentum, and $R$ is the spherical coordinate radius. The physical interest of $dq/dr$ is related to the Rayleigh centrifugal instability. First described by Rayleigh \citep{InstabilitiesCriteria0} for Newtoninan incompressible, and inviscid fluids under axially symmetric rotation, and further generalized to  relativistic flows by \cite{InstabilitiesCriteria}, the instability criterion states that if 
\begin{equation}
    dq/dr<0, 
\end{equation}
then the flow is unstable to small displacements of fluid elements across different layers, which causes convective overturn to develop. 
In Figure~\ref{fig:Rotation1}, we examine the behavior of $dq/dr$ at the equatorial plane of the star.
We note that the only model to which this quantity takes negative values in the PNS's core is Bt0p1e12.
The negative sign of $dq/dr$ of this model is associated with a local minimum of $\Omega$ at $r\sim 7$~km.

The fact that AIC remnants transition to solid rotation was already reported in \citet{Kuroda:2025iyj}. In this earlier work, all their 3D general relativistic models reach a rotation period of $\Omega \sim 4$ kHz, in agreement with our findings. Interestingly, their only magnetized model has an initial poloidal $\vec{B}$-field of $10^{11}$G, comparable to our Bt0p1e11. We argue that (if our results hold for 3D), Bt0p1e11’s core would not be susceptible to MRI instabilities for very long, as it rapidly achieves solid rotation. As a general rule, stronger magnetic fields  make the PNS transition into a solid rotator faster, due to stronger magnetic braking, as we see in Figure~\ref{fig:Rotation1}. Nevertheless, as the Bt0p1e12 core becomes convectively unstable, it holds a higher value of differential rotation at 200ms after bounce. This implies that Bt0p1e12 should face stronger MRI effects. This understanding is in agreement with the fact that \citet{combi2025jet} did manage to resolve MRI instability for a WD progenitor with a poloidal magnetic field of $10^{12}$~G.

\begin{figure*}
    \centering
    \includegraphics[width=1.0\linewidth]{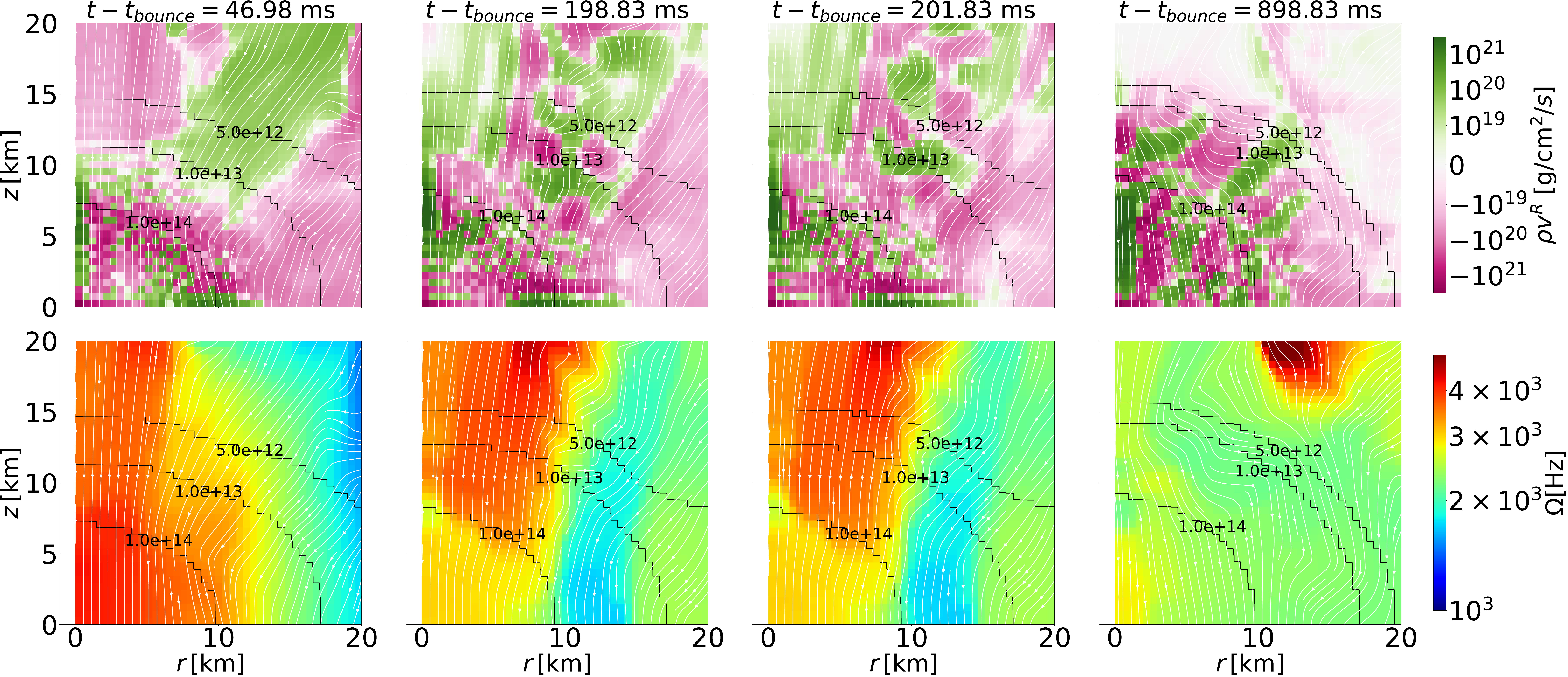}
    \caption{Spherical radial momentum density (top row) and angular velocity (bottom row) profiles for the Bt0p1e12 profile. The white lines represent the poloidal magnetic field, while the black lines represent iso-density contours. Focusing on four time stamps, we see that when the Rayleigh criterion is violated (middle panels), we find a strong convective motion in the outer layers of the star, marked by the intermittent behavior of the sign of $\rho v^{R}$ for increasing R. For earlier and later times, convection is suppressed.    }
    \label{fig:Convection}
\end{figure*}

Figure~\ref{fig:Rotation2} repeats the same analysis but focusing on Bt0p1e12. We can see that the instability sets in after 200~ms post-bounce, only to be quenched 400~ms later. Further evidence for this convective instability is shown in Figure~\ref{fig:Convection}. We find that, when the Rayleigh’s criteria is violated at $t-t_{bounce}\sim$ 200 ms, a stronger convective motion takes place in the outer layers of the magnetar’s core.
It is worth noting that the Rayleigh instability taking place for Bt0p1e12 does not seem to affect the late GW signal (see Figure~\ref{fig:GW}), as it takes place at later times, when the GW emission has already subsided. Maybe this could affect the late-time emission of the $m=1$
instability mode \citep{2023MNRAS.525.6359L}. More 3D studies are necessary for this assessment.

\begin{figure*}
    \centering
    \includegraphics[width=1.0\linewidth]{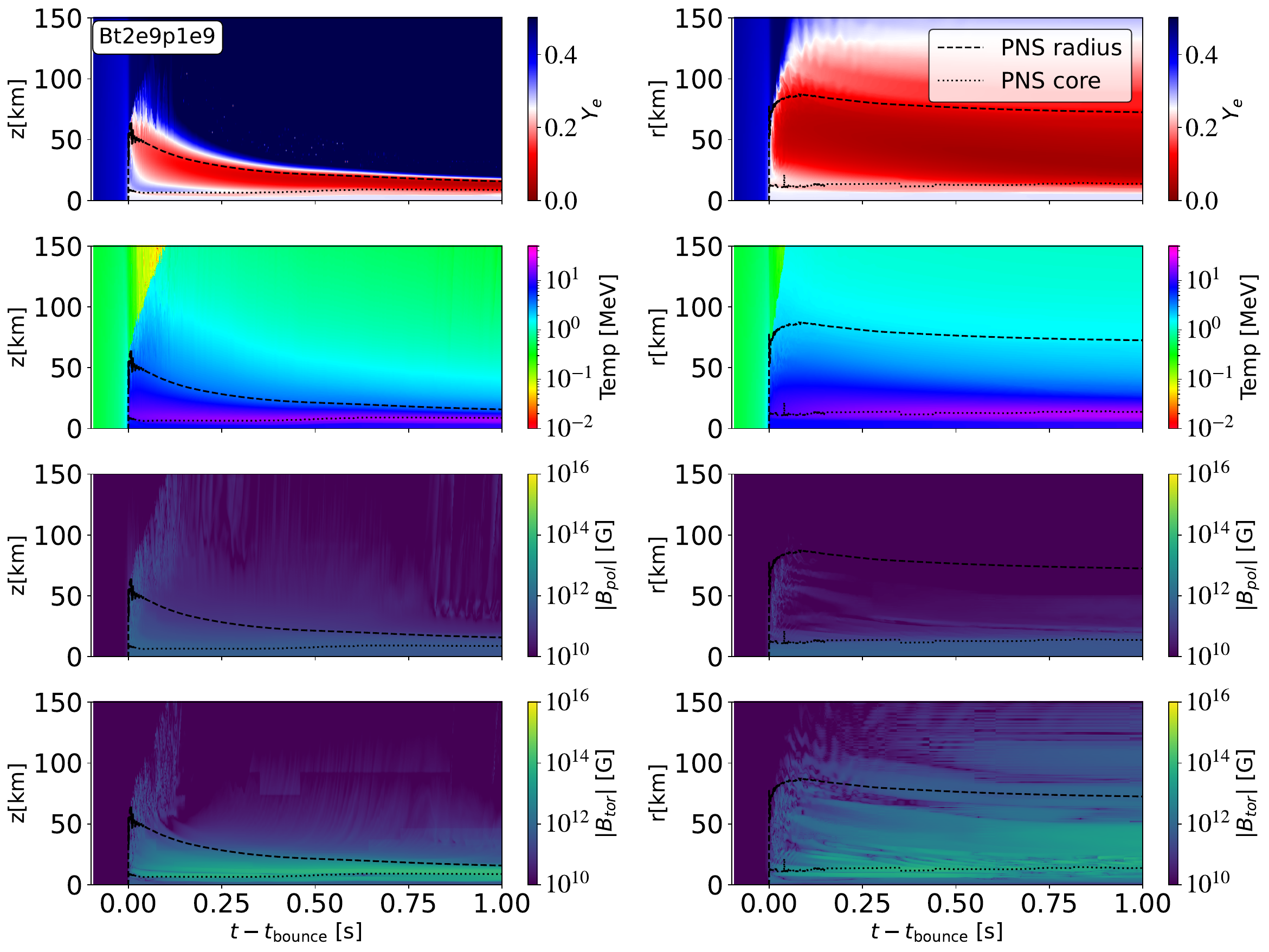}

    \caption{Temporal evolution of the radial profile of the Bt2e9p1e9, as representative of the lower magnetized models. The electron fraction ($Y_e$), the temperature, the poloidal and toroidal field amplitudes ($|B_{pol}|$ and $|B_{tor}|$) are shown as a function of radius(for $z=0$) and of z (for $r=1$ km, as to avoid the coordinate singularity). While a higher $Y_{e}$ is found at central regions of the remnant, the higher temperature is found to be off-center, roughly aligned with the position of the maximum of $B_{\rm tor}$. The black dashed(dotted) line represents the $\rho = 10^{11}(10^{14})$g/cm$^{3}$, representing the estimated surface (core) of the remnant PNS. At $\sim1s$ after bounce, the remnant is still very extended as $r\sim75$km. Most of the magnetic field strength is contained within the magnetar. } 
   \label{fig:Bt2e9remnant}
\end{figure*}

Figure \ref{fig:Bt2e9remnant} shows that, while most of the interior of the PNS is neutron-rich ($Y_{e}\lesssim0.1$), its very core has a relatively higher electron fraction of $\sim 0.25$. This increase of $Y_{e}$ towards the center is due to the neutrino trapping taking place in the center of the remnant.
The remnant also retains temperatures of $\gtrsim 10$ MeV on its core’s surface. The position of the temperature peak aligns with the peak of the magnetic field strength. We understand this correlation by comparing these results against Figure~\ref{fig:Rotation1}. The solid rotation takes place only for the core of the star, which is contained in the $r\lesssim10$ km region. The peak of the temperature and magnetic field takes place on the boundary of this region, where a lot of friction is expected to take place, causing heating and turbulence. These properties create the ideal ambient for high temperature and magnetic field amplification.
Figure \ref{fig:Bt2e9remnant} also shows that the poloidal magnetic field component is well contained within the magnetar’s core. In contrast, the toroidal component extends beyond the extended equatorial radius of the PNS, while maintaining its peak ($\sim 10^{14}$ G) on the core’s surface.

Figure \ref{fig:Bt2e9remnant} shows that, while most of the interior of the PNS is neutron rich ($Y_{e}\lesssim0.1$), its very core has a relatively higher electron fraction of $\sim 0.25$. This increase of $Y_{e}$ towards the center is due to the neutrino trapping taking place in the center of the remnant.
The remnant also retains temperatures of $\gtrsim 10$ MeV on its core's surface. The position of the temperature peak aligns with the peak of the magnetic field strength. We understand this correlation by comparing these results against Figure~\ref{fig:Rotation1}. The solid rotation takes place only for the core of the star, which is contained in the $r\lesssim10$ km region. The peak of the temperature and magnetic field take place on the boundary of this region, therefore a lot of friction is expected to take place, causing heating and turbulence. These properties create the ideal ambient for high temperature and magnetic field amplification. 
Figure \ref{fig:Bt2e9remnant} also shows that the poloidal magnetic field component is well contained within the magnetar's core. In contrast, the toroidal component extends beyond the extended equatorial radius of the PNS, while maintaining its peak ($\sim 10^{14}$ G) on the core's surface.

Figure \ref{fig:Bp1e12remnant} shows the same quantities as Figure \ref{fig:Bt2e9remnant} but for the outlier model Bt0p1e12. While the conclusions about composition and temperature are similar to the ones for Bt2e9p1e9, big differences are seen for the magnetic field configuration.
Different from the other models, the poloidal magnetic field of Bt0p1e12 is not contained within the magnetar. The toroidal component inside the star is also much more turbulent, with regions of $|B_{tol}|\sim 10^{16}$ G away from the core region.

\begin{figure*}
    \centering
    \includegraphics[width=1.0\linewidth]{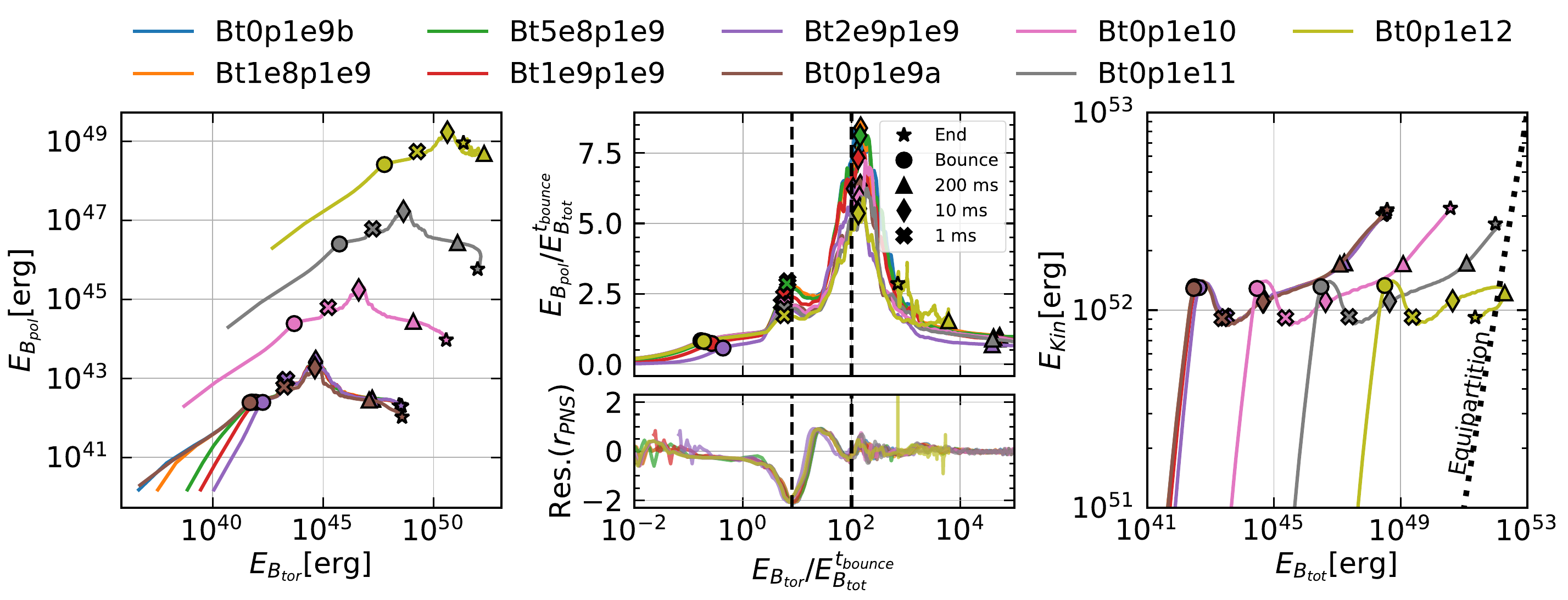}
    \caption{Energy evolution of the PNS interior ($\rho > 10^{11} \mathrm{g/cm^{-3}}$). In the left panel, we see that the magnetic energy configurations at bounce seem to follow a quasi-linear relation between the toroidal and poloidal field energy (in log-space). The middle panel shows the PNS energy evolution but with a normalization by the full magnetic energy at the time of bounce ($E_{B_{\mathrm{tot}}}^{t_{\mathrm{bounce}}}$). The PNS evolution exhibits normalizable profiles across models. Most models follow a similar configuration for all post-bounce phases, except Bt0p1e12. The bottom caption of the middle panels shows the $\rm{Res}(r_{\rm PNS})$, defined at Equation \ref{eq:residual}. The fact that the local minima of these residuals (marking contraction periods) indicate that the post-bounce periodic amplification periods of the poloidal field are due to the oscillations of the PNS. The right caption shows that the only model in which the total magnetic energy surpasses the total kinetic energy is Bt0p1e12 at $t-t_{bounce}\sim200$ms, which we point to as the source of PNS instability.}
    \label{fig:EnergySpace1}
\end{figure*}

In Figure \ref{fig:EnergySpace1}, we perform different analyses on the evolution of the magnetic and kinetic energies of the remnant. Defining the kinetic energy as
\begin{figure*}
    \centering
    \includegraphics[width=1.0\linewidth]{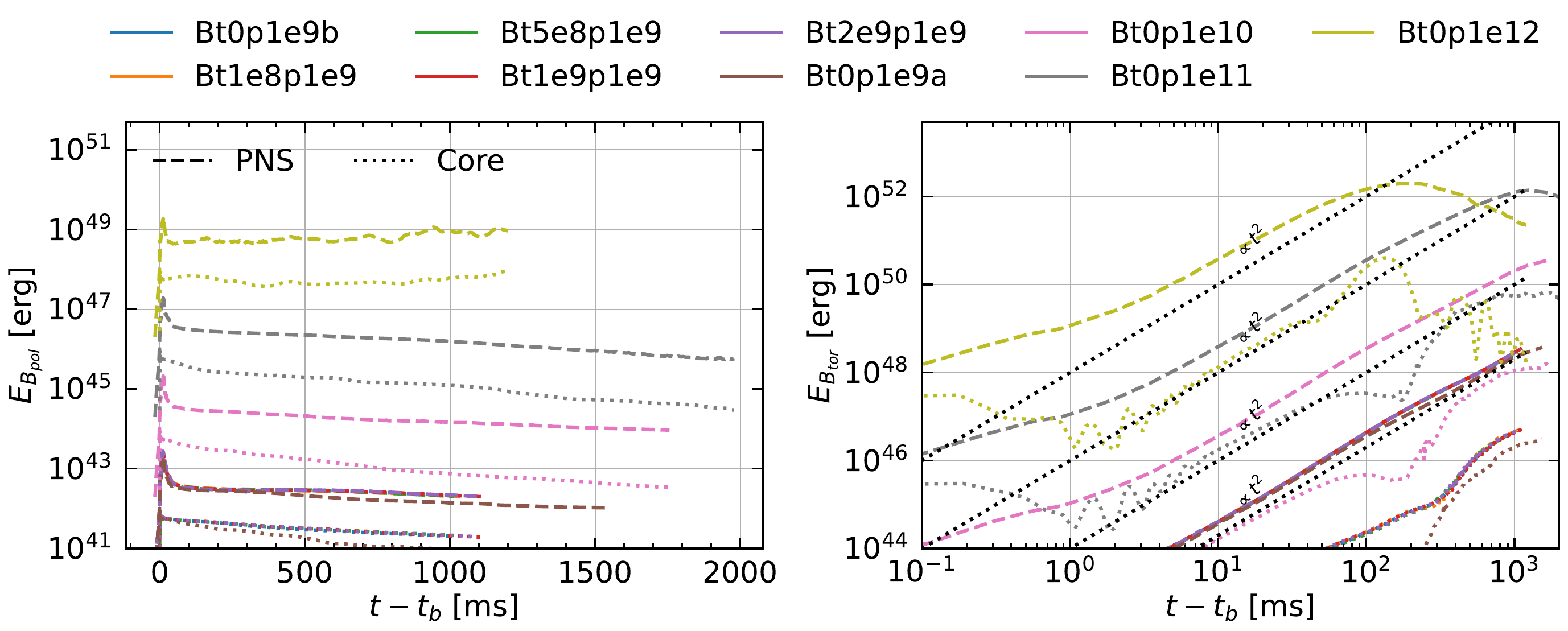}
    \caption{Magnetic field’s energy and maximum value for each of its components. Solid, dashed, and dotted lines display the results for the full grid, $\rho>10^{11}$g/cm$^{3}$, and $\rho>10^{14}$g/cm$^{3}$ integration regions, respectively. The final total energy of the magnetic field is dominated by its toroidal component, which is determined by the progenitor’s $B_{\rm pol}$. This fact is only reinforced by the $\propto t^{2}$ scaling of the toroidal component, which implies growth by winding of the poloidal field component. The exception to this rule is the late time behavior of Bt0p1e12. Bt0p1e12’s PNS (especially its core) shows a strong attenuation of the toroidal magnetic field energy starting at $t-t_{bounce}\sim 200$~ms.}
    \label{fig:Benergy}
\end{figure*}
\begin{equation}
    E_{kin} = \int h W(W-1)\rho dV,
\end{equation}
it includes both rotation energy and radial motion energy, and where $h$ is the fluid specific enthalpy. As we are constraining the volume integral to the PNS interior (or core) densities, after bounce time, the contribution of radial oscillations to the kinetic energy should be negligible, and the latter should be dominated by the rotational energy. We see a clear relation between the energy of the toroidal and poloidal magnetic fields at the time of the bounce. We provide fits of this relation, including and excluding Bt0p1e12, as it represents an outlier, in the sense of being the only one to present an instability and a jet \citep{Cheong_2025_jet}. We fit the relation of the magnetic energies, at bounce time, as
\begin{equation}
    \log_{10}(E_{pol}) = a \times \log_{10}(E_{tor}) + b.
\end{equation}
While including all models leads to $a=1.02 , b=0.301 $ with a root mean square error (RMSE) of 0.166. The exclusion of the outlier model has a negligible impact on the fit, leading to $a=1.03 , b=0.704$ with RMSE of 0.176. Provided that the amplification of the magnetic field components due to flux freezing implies the magnetic field components scale as $|B_{pol}|\propto \rho^{2/3}$ and $|B_{tor}|\propto \rho^{1/3}$. This would imply $a=2$. Therefore, our results show that flux freezing is not the dominant source of magnetic field amplification during collapse. In fact, due to magnetic winding, we expect $|B_{tor}|\propto  ||B_{pol}| \Omega t $, justifying the scaling law that we find. For our dataset, we found that the initial toroidal field is of little influence on the position of the model in the $E_{B_{tor}} \times E_{B_{pol}}$ plane. Further raising the evidence for the memory erasing effect regarding the initial toroidal field energy.

To further examine this relation, we redo the $E_{B_{tor}} \times E_{B_{pol}}$ plot, this time normalized by the value of the PNS total magnetic energy at the time of bounce ($E_{B_{tot}}^{t_{\rm bounce}}$). These results are shown in the middle panel of Figure~\ref{fig:EnergySpace1}. The PNS interior ($\rho > 10^{11}$g/cm$^3$) has a normalizable evolution for all of our models. We note that all models seem to have reached the same point at 200~ms post-bounce, with the notable exception of Bt0p1e12, as it deviates from the normalizable evolution. Notably, this model presents the lowest $E_{B_{pol}}/E_{B_{tot}}^{t_{\rm bounce}}$ ratios among the models. This indicates that poloidal fields are an ubiquitous feature of AIC’s magnetars. Although more simulations are needed to better determine the precise critical value, our data suggest that $E_{B_{pol}}/E_{B_{tot}} \lesssim 6$ at bounce time may serve as an early diagnosis of the convective Rayleigh instability for the core of an AIC’s remnant.

In this post-bounce normalizable evolution of the magnetic energies, we found an oscillatory behavior of $E_{B_{pol}}$ for all of our models. A natural source of amplification for poloidal fields is the conservation of magnetic flux. To claim this cause-effect relation, we analyze the equatorial radius of the PNS's surface ($\rho = 10^{11}$ g/cm$^{3}$), denoted as $r_{\rm PNS}$. The residual of the PNS's surface radial coordinate ($\rm{Res}{(r_{\rm PNS})}$) is shown as an inset at the middle panel of Figure~\ref{fig:EnergySpace1}. $\rm{Res}{(r)}$ is defined as
\begin{equation}\label{eq:residual}
    \rm{Res}{(r_{\rm PNS})} = \bar{r}_{\rm PNS}^{N}(t) - r_{\rm PNS}(t) ,
\end{equation}
where by $\bar{r}^{N}$ we denote the N-points moving time-average of $r_{\rm PNS}$. This quantity is defined to serve as a diagnosis for contraction ($\rm{Res}{(r_{\rm PNS})}<0$) and expansions ($\rm{Res}{(r_{\rm PNS})}>0$) suffered by the PNS. We find that the periods of contractions are roughly aligned with periods of maximum excitation of the poloidal field, corroborating our hypothesis. 
Perhaps the most important effect of these oscillations is the remaining poloidal magnetic field by the end of our simulations.
All the models seem to retain $E_{B_{pol}}/E_{B_{tot}}^{t_{\rm bounce}} \sim 0.5-1.0$ at the end of our simulations, even Bt0p1e12.
When the PNS oscillations cease, the remnant radius is smaller than at the beginning of said vibrations. 
By flux conservation, this amplifies any remaining poloidal field, therefore no model achieves a purely toroidal field by the end of our simulations.

The right and final panel of Figure~\ref{fig:EnergySpace1} shows the kinetic energy as a function of the total magnetic energy. We clearly see that the only model that the PNS configuration crosses to the $E_{B_{tot}}> E_{kin}$ region, even though briefly, is Bt0p1e12 at $t-t_{bounce}\sim200$ ms.
Therefore, we are led to the conclusion that, once the kinetic energy ceases to be dominant with respect to the remnant’s kinetic energy, the PNS becomes convectively unstable, allowing for jet launching and enhanced mass ejection. Given that the toroidal magnetic field is dominant for our models by at least 3 orders of magnitude, see Figure~\ref{fig:Benergy}, our results suggest that the rotational energy of the star is a (zeroth order) upper limit for the toroidal field energy. This instability also allows the field to break out from the PNS interior, as discussed in Figure~\ref{fig:Bp1e12remnant}.

\begin{figure*}
    \centering
     \includegraphics[width=1.0\linewidth]{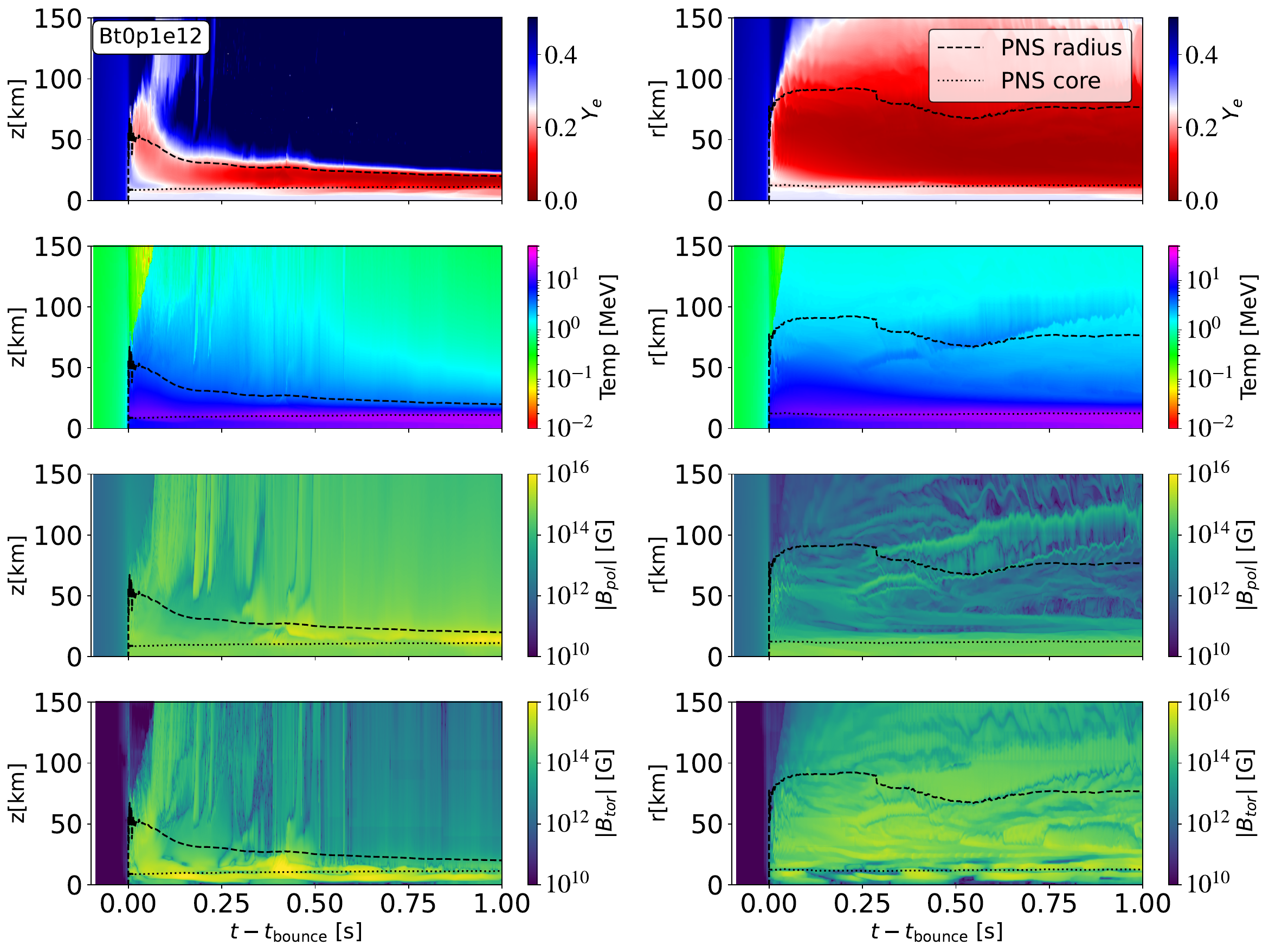}
    \caption{Same as in figure \ref{fig:Bt2e9remnant} but for model Bt0p1e12. We note that the ejection is enhanced at $t-t_{bounce}\sim$ 0.25 ms when the magnetic energy surpasses its kinetic counterpart. At this point, it is clear that the ejected material removes magnetic energy from the remnant, explaining the abrupt decays in $E_{B_{tor}}$ seen in figure \ref{fig:Benergy}. This is manifested as a magnetar flare that propagates outwards from the PNS, seen as the outwards pulse of $|B_{pol}|\sim 10^{14}$G leaving from the PNS's surface $\sim$ 0.3 seconds after bounce.}
    \label{fig:Bp1e12remnant}
\end{figure*}

\subsection{Magnetic Flare Emission}

One of our models (Bt0p1e12) has recently been shown to power a relativistic jet while retaining a spin-down time of the order of ${\sim}10$ seconds, reinforcing a possible connection between AIC and long GRBs with an associated KN \citep{Cheong_2025_jet}. This tentative association is reinforced by further investigations in 3D simulations \citep{combi2025jet}, and by the direct comparison regarding AIC light curves \citep{Pitik:2026bjm}. We report that, prior to the jet launching, a magnetic flare is ejected from the vicinity of the pole, but its energy is rapidly dissipated onto the ejecta, not propagating towards a distant observer.

\begin{figure*}
    \centering
    \includegraphics[width=1.0\linewidth]{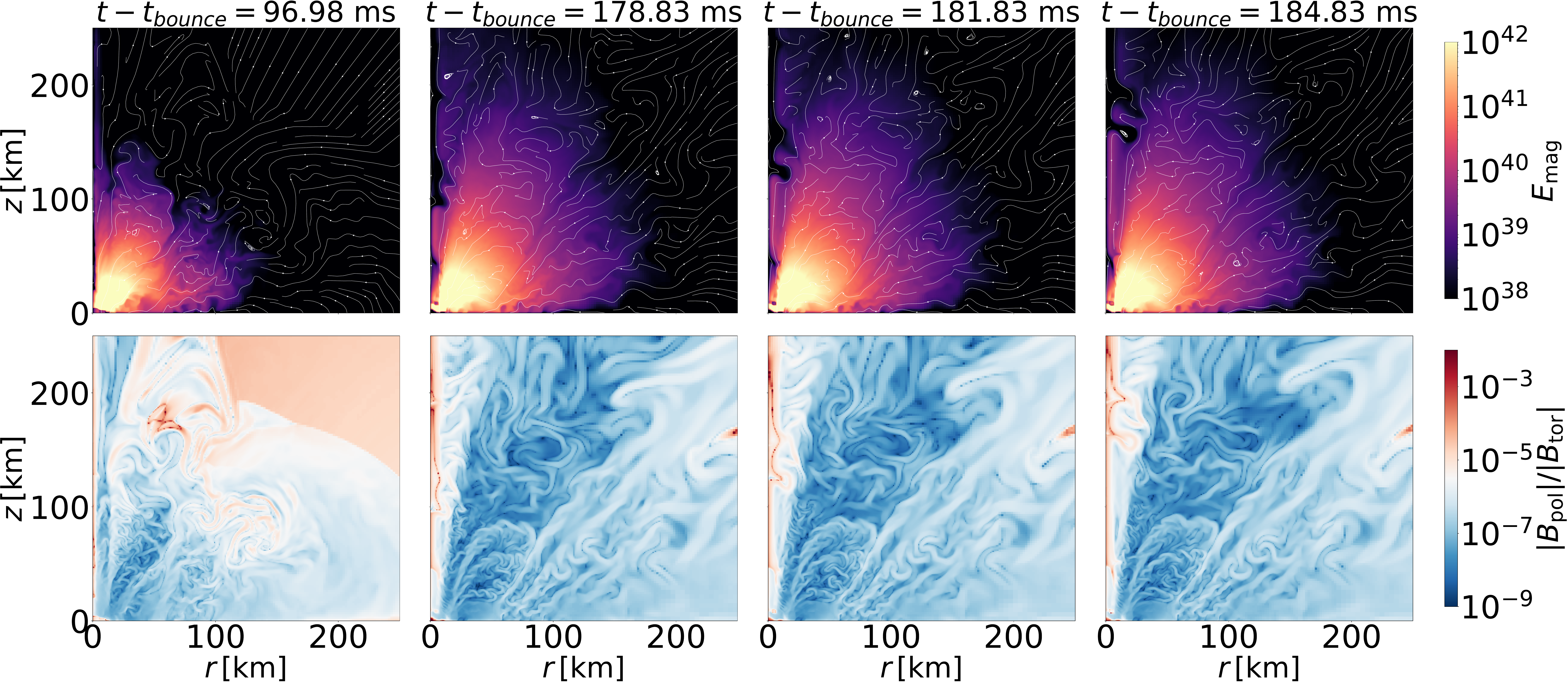}
    \caption{Magnetic energy and poloidal-to-toroidal ratio profiles of Bt0p1e12. The vector field depicted in the image represents the poloidal component of the $\vec{B}$ field. Here we see pockets of magnetic energy leaving the central region, specially on the polar region. The second row of panels show that most of the magnetic field energy is ejected in the form of a toroidal field.}
    \label{fig:flare1}
\end{figure*}

Figure~\ref{fig:flare1} shows snapshots of the outwards propagation of electro-magnetic (EM) energy. At the poles, it is possible to see an upward displacement of clouds of EM energy. In particular, one can trace this more clearly by following the morphology of the magnetic field energy. At the bounce time, most of the surrounding environment display relatively strong poloidal field structure ($|B_{pol}|/|B_{tol}|\gtrsim 10^{-5}$), that is pushed away by the shock that has $|B_{pol}|/|B_{tol}|\lesssim 10^{-6}$. Pockets of relatively high poloidal excitations ($|B_{pol}|/|B_{tol}|\sim 10^{-3}$) form behind the shock and propagate outwards. These pockets could propagate as magnetic flares, as they can, in principle, transport energy to a distant observer. One can also notice the propagation of a magnetic flare in Figure~\ref{fig:Bp1e12remnant}. Noting the pulse of $|B_{pol}|\sim 10^{14}$ G propagating outwards, starting from the PNS’ surface at ${\sim}0.25$~seconds after bounce.

The fact that these pockets of magnetic energy are dominated by toroidal field is in agreement with results shown in Figures~\ref{fig:Benergy} and Figures~\ref{fig:EnergySpace1}, where we see that after $t-t_{bounce} \sim$ 200 ms the toroidal field energy rapidly evolves, diminishing by an order of magnitude in ${\sim} 800$~ms. In the same timescale, the poloidal magnetic field does not show such a rapid monotonic decay. This opens the question regarding the final fate of this ejected energy.

In Figure~\ref{fig:flareenergy}, we compute the electromagnetic luminosity as given by the Poynting vector 
\begin{equation}
\vec{S} = \vec{E} \times \vec{B}= \left( |\vec{B}|^{2}\vec{v} - (\vec{v}\cdot\vec{B})\vec{B} \right),
\end{equation}
where the ideal MHD assumption ($\vec{E} = - \vec{v} \times \vec{B} $) is applied to arrive to the right-hand side. 

Extracting and integrating the flux at consecutively larger cylinders, we find the EM luminosities and energy crossing the cylinders' surface. 
The results shown in Figure~\ref{fig:flare1} demonstrate that the Poynting flux associated with the flare is converted into kinetic/internal energy of the plasma.
For Bt0p1e12, when extracting the EM luminosity at $r=1000$~km, these two activity peaks appear as $\sim10^{51}$erg/s luminosity periods at ${\sim}400$ and ${\sim}800$~ms after bounce. 
These EM flares propagate outwards with $v/c \sim 0.1$, evidenced by the time translation of the peaks in the luminosity profile.
Despite this outward propagation, we find that the total emitted EM energy is dissipated as it propagates away from the remnant.
We realize this by computing the flux integral at larger radii and noting that the total emitted energy drops by orders of magnitude and monotonically decreases as we increase the extraction radius up to 1800 km. In Figure~\ref{fig:flareenergy}, we also include mode Bt2e9p1e9 data for a comparison with a representative model of the remaining models. We show that there is a factor of $10^6$ connecting these models' luminosities. The total magnetic energy reaching 1800km for Bt0p1e12 is of $\sim 3\times10^{50}$ ergs.

\section{ \label{sec:conclusions} Discussions and conclusions}

In this work, by means of the \texttt{Gmunu} code~\citep{2020CQGra..37n5015C, 2021MNRAS.508.2279C, 2022ApJS..261...22C, 2023ApJS..267...38C, 2024ApJS..272....9N}, we perform nine AIC simulations of a fast-rotating WD progenitor ($\Omega \sim 5$ Hz). Our models differ mainly by the initial magnetic field strength and topology. One of our models (Bt0p1e12) is taken as an outlier as it is the only model to launch a (mildly) relativistic jet, see \cite{Cheong_2025_jet}.

Our simulations support the idea that the initial toroidal magnetic field component plays little or no role in the final configuration of the magnetar, which we have described as a memory erasing effect regarding the initial toroidal field. This is shown by the very similar evolution of the PNS remnant in different channels, even when varying the poloidal to toroidal magnetic field excitation. In particular, we found a homogeneous evolution of the poloidal and toroidal field energies when normalized by the total magnetic energy at bounce time.
As the kinetic energy of the remnants is dominant for most of the remnants, the oscillations of the poloidal field are caused by still ongoing vibrations of the PNS. The observed temporal correlation between local maxima of the internal poloidal field and local minima of the PNS radius is consistent with a causal relationship mediated by magnetic flux conservation.
We also show how the magnetic field is mostly constrained inside the remnant, especially for the poloidal component.

\begin{figure*}
    \centering
    \includegraphics[width=1.0\linewidth]{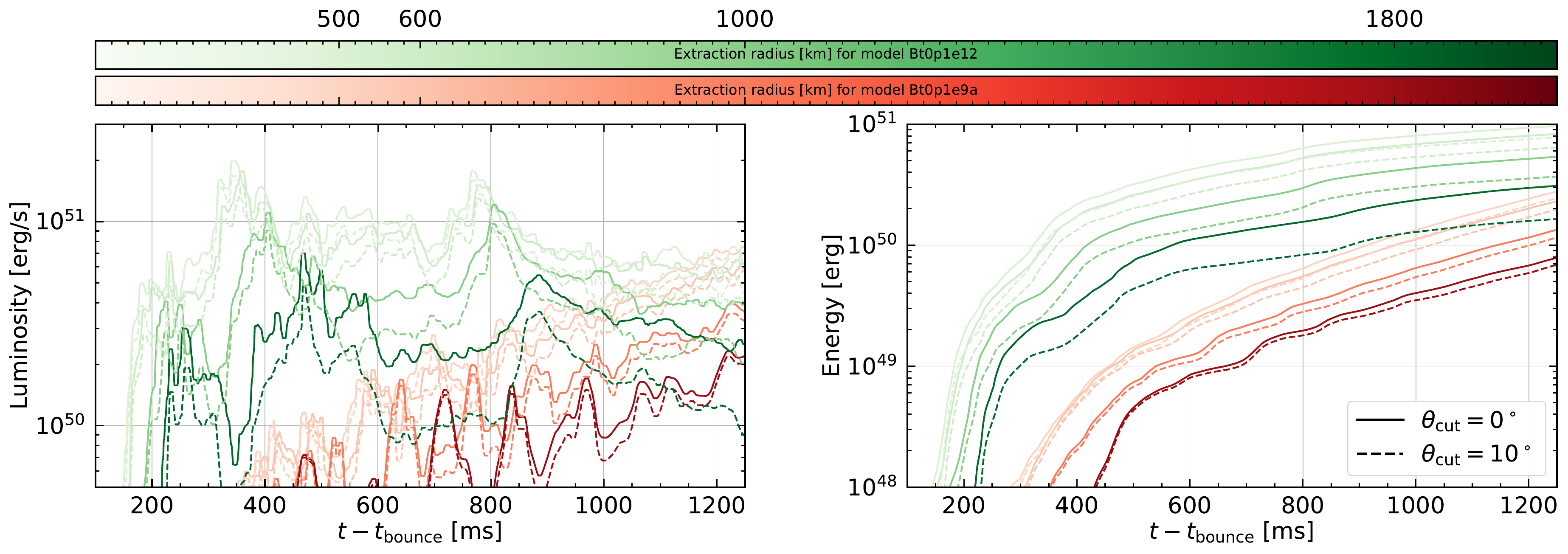}
    \caption{Here we show the magnetic luminosity (left panels) and the integrated released magnetic energy (right panel) for the Bt0p1e9a times $10^6$(red lines) and Bt0p1e12 (green lines) models. We show the results as obtained at different extraction radii. While the magnetic flux in the vicinity of the Bt0p1e12’s PNS ($r\sim$500 km) can exceed $10^{51}$erg/s, the EM luminosity decays rapidly with radius. As the integrated flux decreases with increasing extraction radius (exterior to the PNS), this suggests magnetic energy dissipation via thermalization. For means of comparison, we show that Bt0p1e9a, as representative of the other models, is $10^{6}$ fainter in the EM band. We also performed the same analysis but excluding the Poynting flux through the polar cap ($\theta \lesssim 10^{o}$) to exclude possible contaminations coming from the jet of Bt0p1e12, see \citet{Cheong_2025_jet}. }
    \label{fig:flareenergy}
\end{figure*}

We found an AIC remnant’s core to be stable as long as its kinetic energy is larger than its magnetic counterpart. The only model that temporarily violates this stability condition is Bt0p1e12, starting at $t-t_{bounce}\sim200$~ms. When the PNS’ core becomes unstable, a more violent ejection takes place, carrying both rotational and EM energies. Nevertheless, the rotational energy remains above $3\times 10^{51}$~erg decreasing by ${\sim} 50\%$. The effect on the PNS EM energy is more intense, as it decays from $\gtrsim 10^{52}$~erg to $\lesssim 10^{50}$~erg. Most of this energy comes from the toroidal component of the magnetic field, as shown in Figures~\ref{fig:Benergy} and~\ref{fig:flare1}. The fate of this energy is discussed in Figure~\ref{fig:flareenergy}, where we show that the released EM energy does not reach spatial infinity (at least not in its totality), but it thermalizes very rapidly as it moves to larger radii.
Nevertheless, it is tantalizing to compare these results with GRB’s precursors’ observations. Catalogs of GRB’s precursors show that their estimated isotropic energy can vary from $10^{50}$ to $10^{52}$ ergs for long GRBs \citep{Burlon_2008}. More recently, the isotropic energy of GRB 211211A’s precursor has been estimated to be of $6.97_{1.29}^{+0.15}\times10^{48}$ ergs, $\sim$ 2 orders of magnitude weaker than the flare energy we report for Bt0p1e12. Given that we find that this energy still dissipates as it expands for the duration of our simulations, its reported value should be regarded as an upper limit. A precise comparison between AIC models and GRB’s precursors is left as future work.

As discussed in previous Sections, Figures~\ref{fig:RemnantJM} and~\ref{fig:Rotation2} show that most of our models achieve solid core rotation very quickly after the bounce, despite their initial differential rotation at bounce. Notable exceptions are Bt5e8p1e9 and Bt0p1e12. Although we find no monotonic behavior of the final angular velocity with the increase of the magnetic field, as a general rule, we find that stars with higher degree of magnetization are best solid rotators.

Our results suggest that a maximum toroidal magnetic field exists, which is proportional to the PNS’s rotational energy.
Our results suggest that the remnant stays in a regime where the magnetic energy is subdominant when compared to the kinetic energy. When this condition is no longer satisfied, the remnant goes through an unstable period, ejecting more mass and magnetic energy until the $E_{kin}>E_{B_{tot}}$ condition is satisfied (see Figure~\ref{fig:Benergy}).

Nevertheless, the total magnetic(kinetic) energy in our models is dominated by the toroidal(rotational) component, justifying our previous statement. Regarding the poloidal magnetic field, we find that a minimum magnetic field seems to remain in our models. This seems to be ${\sim} 0.5\ E_{B_{tot}}^{t_{bounce}}$. We understand this remnant field value to be generated by the radial oscillations faced by the PNS via flux freezing. As the PNS is more compact by the end of these oscillations, any initial poloidal magnetic field seeds go through a net amplification.

Although our results are restricted to axisymmetric models, qualitative results are in agreement with 3D simulations such as \cite{2023MNRAS.525.6359L} and \cite{combi2025jet}, showing the robustness of our results. We expect the most significant impact of increasing the dimensionality of our simulations would be for systems that violate the Rayleigh stability criterion. This is because the differential rotation profile faced by our Bt0p1e12 model is prone to source a magneto-rotational instability (MRI) amplification of the $\vec{B}$ fields, which may affect our results. The GW signal is also expected to be different in 3D due to the $m=1$ instability found in \cite{2023MNRAS.525.6359L}.
The extension of this work to 3D is one of the future directions to be pursued. This may be important as the MRI instability can be of importance, in particular at early stages when the star is still differentially rotating. Here, we investigated an isolated supermassive white dwarf, which is not a likely astrophysical scenario. In Nature, it is more likely that a collapsing WD will still be surrounded by material of its former disk/companion. The impact of this extra mass component is left for future work.

Another possible direction of investigation is the extension of these simulations to longer time-scales. Such a study would allow us to investigate the spin-down properties of our models. Analyzing the trajectory of our models in the $\dot{P} \times P$ diagram (where $P$ is the magnetar period), should allow us to perform a more direct comparison of our models and data of magnetar catalogs, such as \citet{ATNFCalogue}. This avenue should enable a better understanding of the AIC event rates in the observable universe. However, for such a project, the ideal-MHD assumption may be too restrictive, and the use of a resistive MHD code may be necessary.

\begin{acknowledgements}

LFLM acknowledges funding from the EU Horizon under ERC Consolidator Grant, no.~InspiReM-101043372 and is also grateful for useful and constructive discussions with Sebastiano Bernuzzi, William Cook, Maximillian Jacobi, and Tetyana Pitik.
P.C.-K.C. acknowledge support from NSF Grant PHY-2020275 (Network for Neutrinos, Nuclear Astrophysics, and Symmetries (N3AS)).
DR acknowledges support from the U.S.~Department of Energy, Office of Science, Division of Nuclear Physics under Award Number(s) DE-SC0024388, and from the National Science Foundation under Grants PHY-2020275, PHY-2116686, PHY-2407681, and PHY-2512802.

The simulations in this work have been performed on the Expanse cluster at San Diego Supercomputer Centre through allocation PHY230104 and PHY230129 from the Advanced Cyberinfrastructure Coordination Ecosystem: Services \& Support (ACCESS) program~\citep{10.1145/3569951.3597559}, which is supported by National Science Foundation grants \#2138259, \#2138286, \#2138307, \#2137603, and \#2138296. 
Additional simulations were performed on the DFG-funded ARA-Cluster of the Friedrich Schiller University Jena.
This research used resources of the National Energy Research Scientific Computing Center, a DOE Office of Science User Facility supported by the Office of Science of the U.S.~Department of Energy under Contract No.~DE-AC02-05CH11231.

We have modified \texttt{RNS}~\citep{1995ApJ...444..306S} to generate the initial data.
The dynamical simulations of this work were produced by utilizing \texttt{Gmunu}~\citep{2020CQGra..37n5015C, 2021MNRAS.508.2279C, 2022ApJS..261...22C, 2023ApJS..267...38C, 2024ApJS..272....9N}.
Neutrino rates are provided by the \texttt{WeakRates} module in \texttt{WhiskyTHC}~\citep{2022MNRAS.512.1499R}.
The ejecta profile is evolved using the 1D radiation hydrodynamics code \texttt{SNEC}~\citep{Morozova:2015bla, Wu:2021ibi}.
The data from the simulations were post-processed and visualised with 
\texttt{yt} \citep{2011ApJS..192....9T},
\texttt{NumPy} \citep{harris2020array}, 
\texttt{pandas} \citep{reback2020pandas, mckinney-proc-scipy-2010},
\texttt{SciPy} \citep{2020SciPy-NMeth} and
\texttt{Matplotlib} \citep{2007CSE.....9...90H, thomas_a_caswell_2023_7697899}.
\end{acknowledgements}

\bibliography{reference}{}
\bibliographystyle{aasjournal}

\begin{figure*}[!t]
    \centering
    \includegraphics[width=\linewidth]{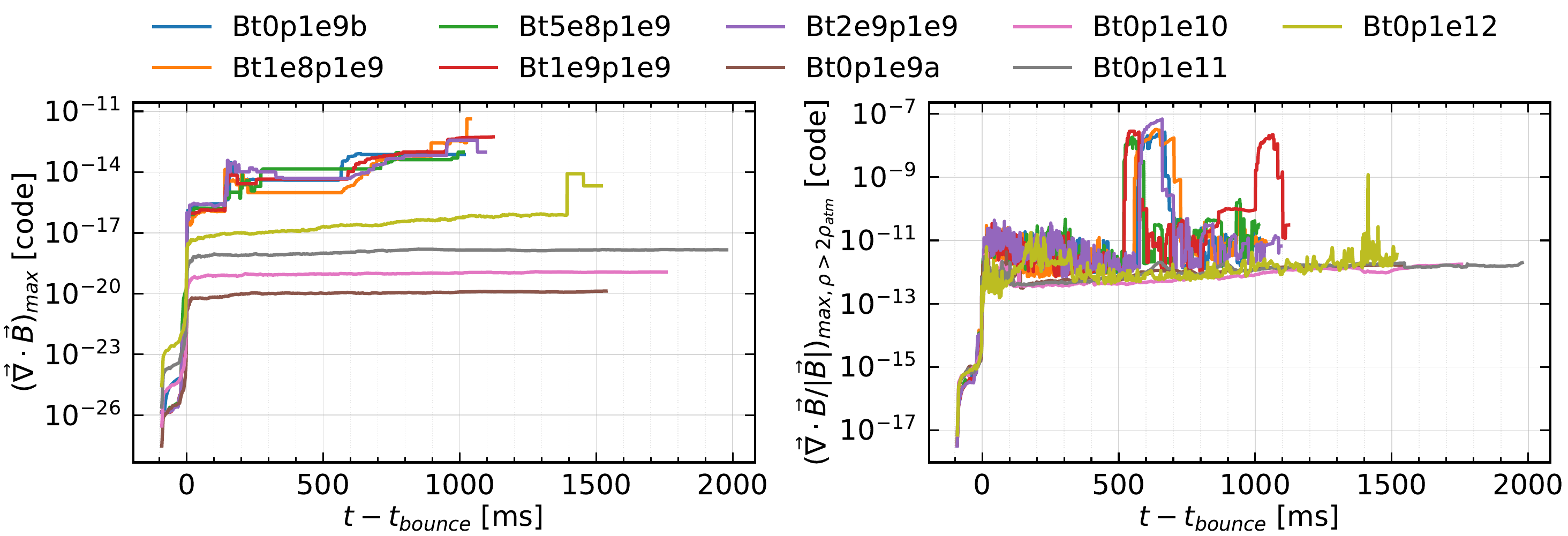}
    \caption{Maximum value of the magnetic field's divergence ($\vec{\nabla}\cdot{\vec{B}}$) for the full grid in all of our models. All models retain a violation below $10^{-11}$. When normalizing the divergence by the magnetic field strength ($\vec{\nabla}\cdot{\vec{B}}/|B|$), we restrict ourselves to regions where $\rho>2\rho_{\mathrm{atm}}$, to avoid including errors due to low atmospheric densities.}
    \label{fig:Constraint}
\end{figure*}

\appendix 
\section{\label{sec:Diagnostics}Diagnostics}
\subsection{Constraint violation}

Figure~\ref{fig:Constraint} shows the maximum constraint violation found in our models. All our models maintain a constraint violation below $\vec{\nabla}\cdot\vec{B}<10^{-11}$ until $\lesssim1$ second. For densities above twice the atmospheric density ($\rho_{\mathrm{atm}}\sim 6\times10^{8}\,\mathrm{g\,cm^{-3}}$), we find $\vec{\nabla}\cdot\vec{B}/|\vec{B}| \lesssim 10^{-11}$ most of the time, with small periods of greater violation ($\lesssim 10^{-7}$). The simulations were stopped shortly after the models violated this threshold.

\end{document}